\documentclass[11pt,letterpaper]{article}
\pdfoutput=1

\usepackage{jcappub}

\usepackage{graphicx}
\usepackage{dcolumn}
\usepackage{bm}
\usepackage{natbib}
\usepackage{amsmath}
\usepackage{amssymb}
\usepackage{hhline}
\usepackage{color}
\usepackage[T1]{fontenc}
\usepackage{wasysym}

\newcommand{\be}{\begin{equation}}
\newcommand{\ee}{\end{equation}}
\newcommand{\bea}{\begin{eqnarray}}
\newcommand{\eea}{\end{eqnarray}}

\newcommand{\Mpc}{{\rm ~Mpc}}

\definecolor{darkgreen}{rgb}{0,0.5,0}

\begin{document}

\title{Spherical Cows in Dark Matter Indirect Detection}

\author[a,b]{Nicol\'{a}s Bernal,}

\affiliation[a]{Centro de Investigaciones, Universidad Antonio Nari\~{n}o\\
Cra 3 Este \# 47A-15, Bogot\'{a}, Colombia}

\affiliation[b]{ICTP South American Institute of Fundamental Research\\
Instituto de F\'{i}sica Te\'{o}rica, Universidade Estadual Paulista\\
R. Dr. Bento Teobaldo Ferraz 271, 01140-070 S\~{a}o Paulo,  Brazil}

\author[c]{Lina Necib,}

\author[c]{and Tracy R. Slatyer}

\affiliation[c]{Center for Theoretical Physics, Massachusetts Institute of Technology, Cambridge, MA 02139, USA}

\emailAdd{nicolas.bernal@uan.edu.co}
\emailAdd{lnecib@mit.edu}
\emailAdd{tslatyer@mit.edu}

\date{today}

\abstract{Dark matter (DM) halos have long been known to be triaxial, but in studies of possible annihilation and decay signals they are often treated as approximately spherical. In this work, we examine the asymmetry of potential indirect detection signals of DM annihilation and decay, exploiting the large statistics of the hydrodynamic simulation Illustris. We carefully investigate the effects of the baryons on the sphericity of annihilation and decay signals for both the case where the observer is at 8.5 kpc from the center of the halo (exemplified in the case of Milky Way-like halos), and for an observer situated well outside the halo. In the case of Galactic signals, we find that both annihilation and decay signals are expected to be quite symmetric, with axis ratios very different from 1 occurring rarely. In the case of extragalactic signals, while decay signals are still preferentially spherical, the axis ratio for annihilation signals has a much flatter distribution, with elongated profiles appearing frequently. Many of these elongated profiles are due to large subhalos and/or recent mergers. Comparing to gamma-ray emission from the Milky Way and X-ray maps of clusters, we find that the gamma-ray background appears less spherical/more elongated than the expected DM signal from the large majority of halos, and the Galactic gamma ray excess appears very spherical, while the X-ray data would be difficult to distinguish from a DM signal by elongation/sphericity measurements alone.}


\preprint{MIT-CTP/4777}
\maketitle

\section{Introduction}

Gravitational evidence for dark matter (DM) is well established \cite{Zwicky:1933gu,Rubin:1970zza,Markevitch:2003at}, yet DM still evades all other means of detection \cite{Akerib:2015rjg,Agnese:2015nto,Ackermann:2015zua,TheFermi-LAT:2015kwa,Khachatryan:2014rra}. A current focus of the search for DM is indirect detection: DM annihilation or decay could produce observable Standard Model (SM) particles, including photons. If such a signal was detected, the direction of the incoming photons could be used to map out the morphology of DM halos. 

N-body simulations constitute a valuable tool for studying the expected DM distribution \cite{1961AJ.....66..590V,1963MNRAS.126..223A,1992ApJ...391..502K,1983ApJ...270..390M,Bryan:2012mw}, and can be used to predict the properties of indirect signals from DM \cite{Stoehr:2003hf,Bernal:2014mmt,Calore:2015oya}.
Hydrodynamic simulations include baryonic matter as well as DM, and thus can probe the impact of baryonic feedback on the DM distribution \cite{Bryan:2012mw}. With recent hydrodynamic simulations that generate large ensembles of DM halos, we can make statistical statements about the general properties of DM halos with and without baryons \cite{Dubinski:1991bm,Warren:1992tr,Dubinski:1993df,Jing:1994sg,Jing:2002np,Kazantzidis:2004vu,Allgood:2005eu,Debattista:2007yz,Schneider:2011ed,Bruderer:2015rnh}. In particular, as we demonstrate in this work, we can map out the full distribution of properties relevant to indirect DM searches, rather than relying on a small number of example halos.

In this article, we focus on studying the morphology of indirect detection signals using N-body simulations. More specifically, we study sphericity/asymmetry of signals after projection along the line of sight. We perform a statistical analysis of the annihilation/decay signatures of a large number of halos in two simulations: the hydrodynamic simulation Illustris-1, which includes DM and baryons, and its DM-only equivalent Illustris-1-Dark \cite{Vogelsberger:2014dza,Vogelsberger:2014kha}. We predict the shape of annihilation/decay DM signals from Galactic and extragalactic (EG) sources, and diagnose the effect of baryons on the asymmetry and sphericity of these signals. For the remainder of the text, we will refer to signals as ``spherical'' if they could be produced by the line-of-sight projection of a spherical 3D source of photons; i.e. they are symmetric under rotation of the sky around the line-of-sight pointing toward their center.

Several potential signals have appeared in indirect DM searches over the past few years. An anomalous emission line at $\sim$3.5 keV has been found in a stacked analysis of 73 galaxy clusters \cite{Bulbul:2014sua} and in other regions \cite{Ruchayskiy:2015onc,Iakubovskyi:2015dna,Bulbul:2014ala,Franse:2016dln}. Analysis of data from the \textit{Fermi} Gamma-Ray Telescope (hereafter \textit{Fermi}) \cite{Atwood:2009ez} has shown an unexplained spherically symmetric excess of $\mathcal{O}$(GeV) gamma rays at the center of the Galaxy \cite{Goodenough:2009gk,Abazajian:2012pn,Daylan:2014rsa,Zhou:2014lva,Calore:2014xka,TheFermi-LAT:2015kwa,Linden:2016rcf}. Studying expected properties could help discriminate DM against astrophysical backgrounds. 

This paper is organized as follows. First we introduce our methodology in Sec.~\ref{sec:illustris}; we describe the Illustris simulation, and the related computations of DM density, and define the metrics used for the determination of halo shapes. We then perform two analyses of annihilation and decay signals, one where the observer is situated at a location 8.5 kpc from the center of the halo (Sec.~\ref{sec:results}), and one where the observer is outside the halo (Sec.~\ref{sec:extragalactic}). In each of these sections, we present the overall distributions for asymmetry and axis ratio.  For the former (Galactic) analysis, we focus on the subcategory of Milky-Way (MW) type halos. For the latter (extragalactic) analysis, we focus on cluster-sized halos. In both cases, we also study possible correlations between halo axes and the baryonic disk. In Sec.~\ref{sec:observations}, we discuss two case studies of the morphology of astrophysical backgrounds for DM searches, first considering the gamma-ray background and signal for the \textit{Fermi} inner galaxy excess, and then the clusters in which the $\sim 3.5$ keV line is detected. We summarize and conclude in Sec.~\ref{sec:conclusion}.

\section{Methodology} \label{sec:illustris}

In the context of indirect DM searches for photons or neutrinos, the quantity of interest for decay (annihilation) is the integrated DM density (density squared) of DM particles along the line of sight. This is referred to as the $J$-factor; to compute it within the Illustris simulation, we must define it in the context of the discrete representation of the underlying matter distribution.

\subsection{Illustris Simulation }
The Illustris simulation is a publicly available\footnote{\url{http://www.illustris-project.org/}} hydrodynamic simulation that traces the evolution of DM particles, as well as gas, stars and black holes across redshifts from $z = 127$ to today $z = 0$ \cite{Vogelsberger:2013eka,Torrey:2013pwa,Vogelsberger:2014dza,Vogelsberger:2014kha,Genel:2014lma}. The Illustris simulation employs a comprehensive suite of baryon physics including stellar evolution and feedback, gas recycling, supermassive black hole growth, and feedback from active galactic nuclei \cite{Vogelsberger:2014dza}. In this work, we focus on the last snapshot at $z= 0 $, which reflects the simulated state of today's Universe \cite{Nelson:2015dga}. The simulation is conducted at 3 different resolution levels, Illustris-1, Illustris-2, and Illustris-3. It also includes the same set of simulations for DM only particles, at the same resolution levels, Illustris-1-Dark, Illustris-2-Dark, and Illustris-3-Dark. The simulations cover a total volume of $(106.5 \Mpc)^3$. In this work, we focus on the highest resolution simulations Illustris-1 and Illustris-1-Dark.  Parameters of the simulations are shown in Table \ref{tab:illustris}, including the mass of the DM and baryon particles, and the spatial resolutions of the simulations. The mass of a DM particle is fixed throughout the simulation, but that of a baryonic particle (which sums the mass of the gas, stars and black holes) is not conserved, but kept within a factor of 2 of the quoted baryonic mass $m_{\rm{b}}$. Gravity is included with softening of the potential at small scales to avoid numerical two-body particle scattering \cite{Springel:2009aa}. 
The softening lengths for both DM and baryons, $\epsilon_{\rm{DM}}$ and $\epsilon_{\rm{b}}$, are also shown in Table \ref{tab:illustris}. 

\begin{table}[t]
\centering
\begin{tabular}{|c||c|c|c|c|}  
\hline 
Simulation & $m_{\rm{DM}} (M_{\astrosun})$ & $m_{\rm{b}} (M_{\astrosun})$ & $\epsilon_{\rm{DM}} ({\rm{kpc}}) $& $ \epsilon_{\rm{b}} ({\rm{kpc}}) $ \\ \hline \hline
  Illustris-1  & $6.3 \times 10^6 $& $1.3 \times 10^6$  & 1.4 & 0.7 \\
  Illustris-1-Dark & $7.5 \times 10^6$ & $-$  & 1.4 & $-$ \\ \hline
\end{tabular} 
\caption{The particle masses and softening lengths for the Illustris-1 and Illustris-1-Dark simulations \cite{illustris}; ``DM'' subscripts label DM, while ``b''  subscripts label baryons.} \label{tab:illustris}
\end{table}

The Illustris simulation used the friends-of-friends (FOF) algorithm to identify DM halos \cite{Davis:1985rj}. The Illustris-1 simulation has 7713601 halos, and 4366546 identified as subhalos. The Illustris-1-Dark simulation includes 4263625 halos and 4872374 subhalos.  To limit the impact of poorly resolved objects on our results, we only examine halos with at least 1000 DM particles. This cut leaves $1.6 \times 10^{5}$ and $1.5 \times 10^{5}$ halos for Illustris-1 and Illustris-1-Dark respectively. The halo mass function of the Illustris simulation described in Ref. \cite{Vogelsberger:2014dza} is in good agreement with the empirical data. Deviations from observations are present at the low and high end of the resolved masses, where the details of the implementation of the stellar and AGN feedback are important. The mass range of halos that pass the 1000 DM particle cut is $\sim 5 \times 10^{9} M_{\astrosun} $ to $3 \times 10^{14} M_{\astrosun}$. 

\subsection{Computing $J$-factors}
\label{sec:jfactor}
The quantity of interest in this analysis is the $J$-factor, defined by 
\be \label{eq:jfactor}
J = 
\begin{cases}
\int \rho^2~ ds~ d\Omega & \text{for annihilation}, \\
\int \rho ~ds~ d\Omega & \text{for decay},
\end{cases} 
\ee
where $\rho$ is the density of DM, and the integral is along the line of sight and over solid angles. In order to integrate the local DM density (or density squared) given the discrete particle distribution, we use a kernel summation interpolant to reconstruct a continuous DM density field \cite{1985A&A...149..135M, Hernquist:1988zk, Springel:2011yw}. More explicitly, for a field $F(\vec{r})$, one can define a smoothed interpolated version $F_s(\vec{r})$, related to $F$ through a kernel function $W$
\be
F_s(\vec{r}) = \int F(\vec{r'}) \, W(\vec{r} - \vec{r'}, d)\, dr'\, ,
\ee
where $d$ defines the length of the smoothing. The kernel function $W$ approaches a delta function as $d \rightarrow 0$. We use a cubic spline to compute the local density $\rho$:
\be
w (q) = \frac{8}{\pi d^3} \begin{cases} 1 - 6 q^2 + 6 q^3 ; & 0 \leq q \leq \frac{1}{2} \\ 
2 (1 - q)^3; & \frac{1}{2} < q \leq 1 \\
0; & q> 1. \end{cases} 
\ee
The smoothing kernel in this case is $W(r,d) = w(r/2d)$, where $d$ is chosen to be the distance to the 33$^{\rm{rd}}$ nearest neighbor of the point considered. In order to compute the density along a particular line of sight defined by the galactic coordinates $(l,b)$, we sum over the density of the 32 nearest neighbors to a particular point, and adjust the next step in the integral to be the newly found $d$. We have checked that doubling the number of neighbor particles employed in this procedure from 32 to 64 does not alter our results. This is due to the fact that the contribution of further particles is proportional to $1/r^2$.

We proceed to construct sky maps of annihilation/decay $J$-factors for each halo, by placing an observer at $R_\odot = 8.5$ kpc from the center of the halo along the $x$-axis of the simulation, and compute the $J$-factor for different values of galactic coordinates $(l,b)$. The center of the halo is defined as the location of the gravitational potential minimum. We use the package HEALPix\footnote{\url{http://healpix.sourceforge.net}} to divide the sky into equal area pixels \cite{2005ApJ...622..759G}. The total number of pixels in a map is defined by 
\be \label{eq:nside}
n_{\rm{pix}} = 12 \times \text{nside}^2,
\ee
where nside is an input parameter that defines the pixelation. 
\begin{figure}[t]
\begin{center}
\includegraphics[trim={0 0 0 1.6cm},clip,width=0.6\textwidth]{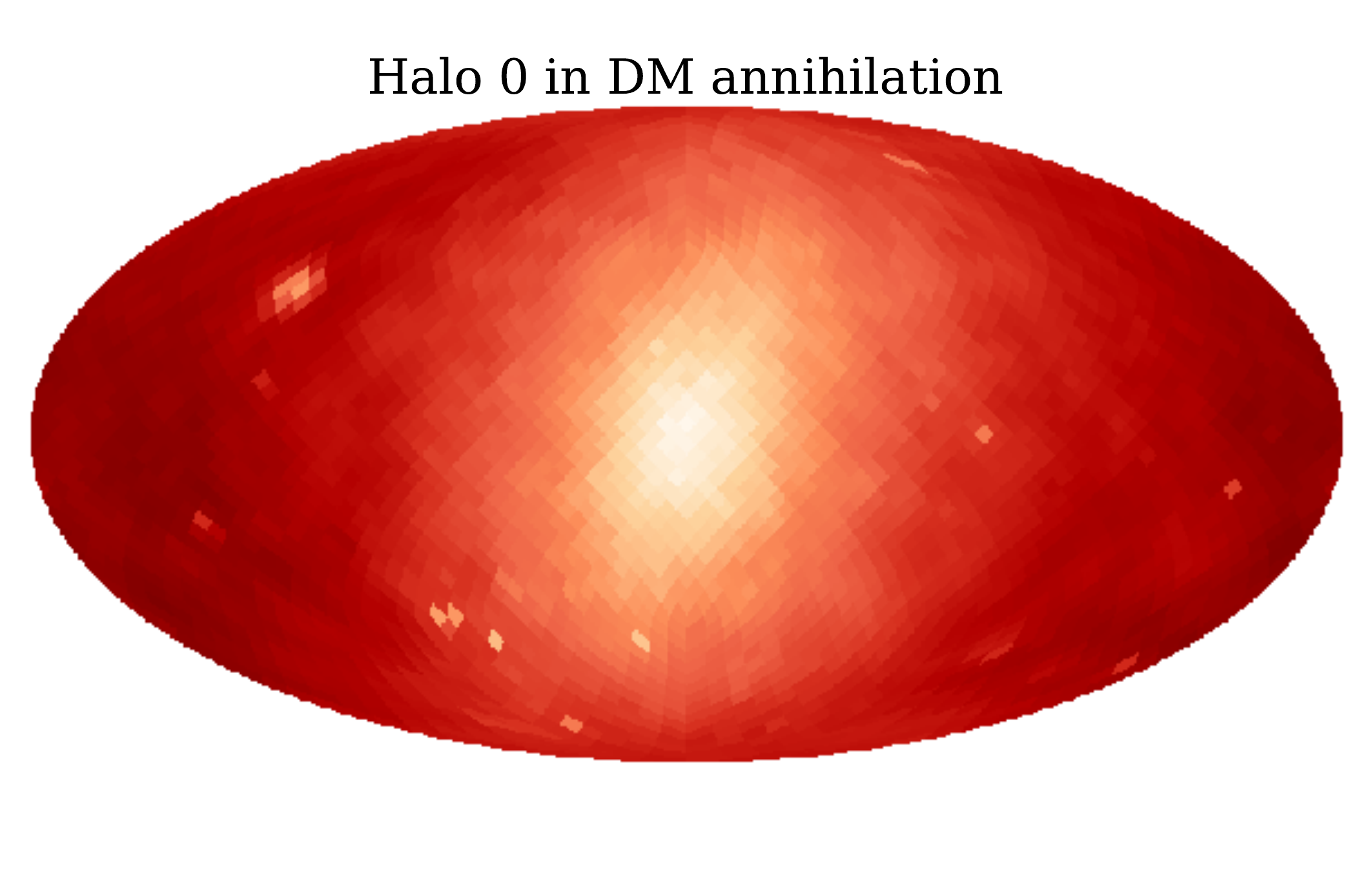}
\vspace{-10 mm}
\end{center}
\caption{\label{fig:maps} Logarithmic map for DM annihilation for halo labeled ``0'' in the Illustris-1 simulation. Lighter colors signify higher DM \emph{luminosity}. We used HEALPix with nside $= 16$ (see Eq. \ref{eq:nside}). This map is taken by positioning the observer at 8.5 kpc from the center of the halo. The mass of this halo is $3.2 \times 10^{14} M_{\astrosun}$. 
}
\end{figure}
We show an example of such constructed maps in Fig. \ref{fig:maps}, for the annihilation signal of a particular halo. For the example chosen, the halo radius, $R_{200}$ defined as the radius such that the average density interior to that radius is 200 times the critical density of the universe,\footnote{The distances in the simulation are presented in units of kpc/$h$, where $h = 0.704$ is the reduced Hubble constant so that $H = h \times 100 $ km/sec/Mpc. The cosmological parameters used are $\Omega_m=  0.2726$, and $\Omega_\Lambda = 0.7274 $ \cite{Vogelsberger:2014dza}.} is 1659 kpc$/h = 2356$ kpc, much larger than the observer distance from the center of the halo $R_\odot  = 8.5$ kpc, and therefore, there is still sizable signal from high latitudes. 

\subsection{Asymmetry Parameterization} 
\subsubsection{Axis Ratio}
\label{sec:axis_ratio}

One measure of sphericity is the axis ratio. First, we summarize previous methods of finding the axis ratio and the major axis. Unlike previous analyses that analyzed the halo shapes through the 3D moment of inertia tensor \cite{1986ApJ...304...15B,1987ApJ...319..575B,Dubinski:1991bm,Jing:2002np,Allgood:2005eu,Schneider:2011ed}, 
we compute the 2D projection of the inertia tensor along the plane perpendicular to the line between the observer and the halo center, as all indirect detection signals are found in projection. That is defined as \cite{Tenneti:2014poa}
\begin{equation} \label{eq:inertiatensor}
 I_{i,j} = \sum_{n} x_{n,i}\, x_{n,j},
\end{equation}
where the sum is taken over the DM particles $n$ of the halo, and $i,j$ correspond to the coordinates of the particle $n$ projected on the plane perpendicular to the observer.\footnote{See Ref. \cite{2012MNRAS.420.3303B} for a discussion of the different definitions of the inertia tensor that occur in the literature. } For example, if the observer is located along the $x$ axis (where the center of the Cartesian coordinate system is at the center of the halo, defined in the simulation as the location of the most-bound particle), $x_i,x_j$ run over the four combinations of the $y$ and $z$ coordinates for each of the DM particles of the halo. The axis ratio is defined as the ratio of the square root of the eigenvalues of the inertia tensor, where in this work we use the convention where the axis ratio is always less than 1. The major axis of the halo is the eigenvector corresponding to the largest eigenvalue. In this notation, the axis ratio of a spherical halo is 1.

We introduce a variation on the inertia tensor defined in Eq. \ref{eq:inertiatensor} that is adaptable to indirect detection signals.  This new inertia tensor uses the same information that we would have looking at a DM annihilation/decay signal. In this case, the DM particle coordinates are weighed by \emph{luminosity} in DM signal, which is the $J$-factor at that location. The new inertia tensor that we call the $\mathcal{J}$-tensor is therefore
\begin{equation} \label{eq:jfactortensor}
 \mathcal{J}_{i,j} = \sum_{n} J(z_{n,i}, z_{n,j})\, z_{n,i}\, z_{n,j},
\end{equation}
where the coordinates $z_{n,i}$ are obtained from scanning through the pixels in the sky and inferring the Cartesian coordinates of that particular pixel (assuming we live in a sphere), and $J$ is given by Eq. \ref{eq:jfactor} at a point in the sky given by the coordinates $z_{n,i}$. With this approach, all particles within the same line of sight contribute only once but their contribution is weighed with the observed intensity of the signal. As above, the observed axis ratio is defined as the ratio of the square roots of the eigenvalues of the $\mathcal{J}$-tensor, and the halo's major axis is the eigenvector with the largest eigenvalue.

\subsubsection{Quadrant Analysis} 
\label{sec:quadrants}

As a second parameterization of the observed asymmetry of DM signals, we divide the observed sky into four equal quadrants, with the origin of the coordinate system lying along the line of sight to the center of the halo. The halos are oriented randomly relative to the quadrant boundaries unless otherwise stated.

We then determine the $J$-factor associated with each quadrant as discussed in Sec. \ref{sec:jfactor}
\begin{align}
 J_k = \sum_{i \in R_k} J_i,
\end{align}
where $k \in \{1,2,3,4 \}$ labels the different quadrants, and $R_k$ is the list of pixels in quadrant $k$. $J_i$ is the value of the $J$-factor found at pixel $i$. The quadrants are labeled such that quadrant 1 is adjacent to 2 and 4, and opposite to 3. 
We define the following ratios, describing the relative predicted emission in pairs of opposite or adjacent quadrants

\begin{align} R_\mathrm{opp} & = \frac{| (J_1 + J_3) -( J_2 + J_4)|}{\sum_i J_i},  \label{eq:radj} \\
 R_\mathrm{adj} & = \frac{|(J_1 + J_2 )- (J_3 + J_4)|}{\sum_i J_i}  .  \label{eq:ropp} \end{align}

For signals that appear spherical from the point of view of the observer, the ratios defined in Eqs. \ref{eq:radj} and \ref{eq:ropp}  $R_{\text{opp}} = R_{\text{adj}} = 0$. For signals that appear strongly elongated or asymmetric, $R_\mathrm{opp} \rightarrow 1$ or $R_\mathrm{adj} \rightarrow 1$  depending on which quadrants dominate the DM signal.

\section{Galactic Analysis}
\label{sec:results}

\begin{figure*}[t]
\begin{center}
\includegraphics[width=0.45\textwidth]{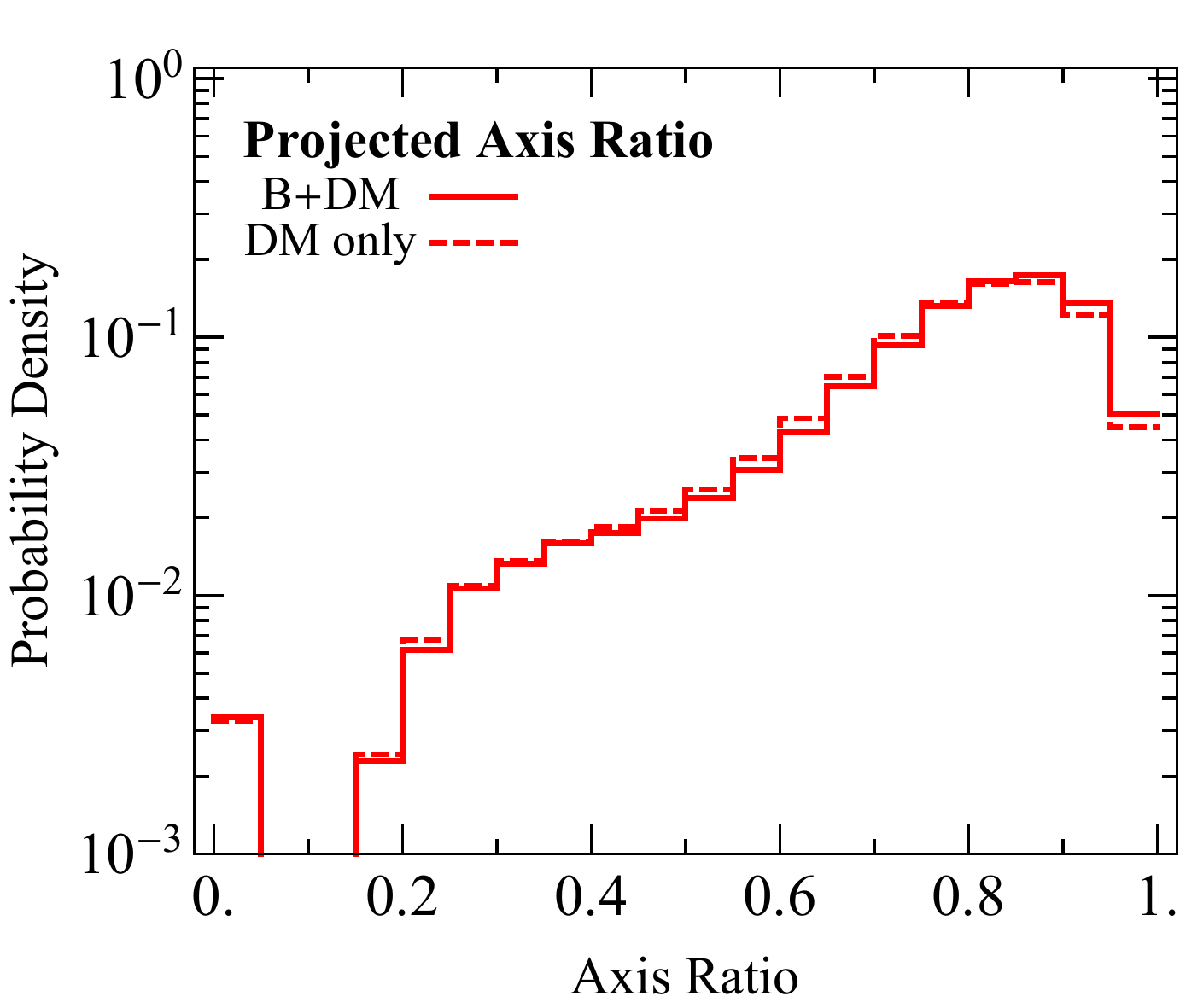}
\includegraphics[width=0.45\textwidth]{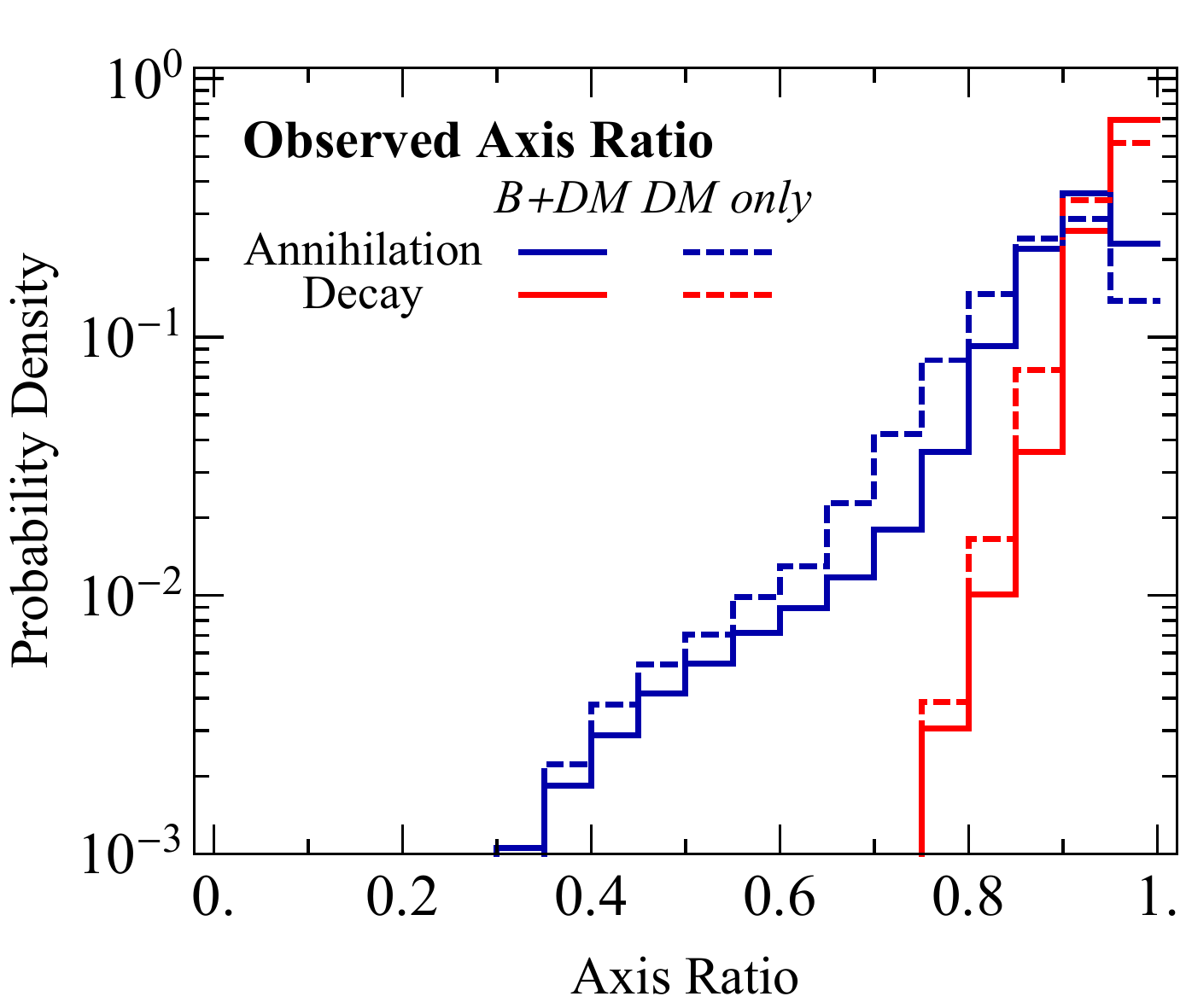}
\caption{\label{fig:axis_ratio}
 Histogram of the conventional 2-dimensional axis ratio (left) and the newly defined observed axis ratio (right) for annihilation and decay, comparing both Illustris-1 and Illustris-1-Dark (see Eq. \ref{eq:jfactortensor}).
}
\end{center}
\end{figure*}

In this section, we perform a statistical analysis of the sphericity of annihilation/decay signals as observed from a location $R_{\odot} = 8.5 $ kpc from the halo center, similar to the Earth's separation from the center of the Milky Way \cite{2009A&A...498...95V, Gillessen:2008qv}. We first study the distribution of the observed axis ratio in annihilation and decay as defined in Sec.~\ref{sec:axis_ratio}, as well as the distribution of the ratio in intensity of opposite and adjacent quadrants as introduced in Sec. \ref{sec:quadrants}. We plot the histograms of probability densities in each distribution, counting each halo just once unless otherwise stated. We then focus our analysis on Milky-Way-like halos, orienting the observer to be on the halo disk, and study the axis ratio distribution. We finally examine the possibility of correlations between the halo minor axis and the baryonic disk. 

\subsection{Observed Axis Ratio}
\label{sec:obs_axis_ratio}

For comparison, we first illustrate the distribution of the axis ratio as obtained from the two-dimensional inertia tensor defined in Eq. \ref{eq:inertiatensor}. As shown in Fig. \ref{fig:axis_ratio} (left), we find that the distribution peaks at axis ratio $\approx 0.85$, which is consistent with results found from the projected shapes of DM halos inferred from the position of galaxies in the Sloan Digital Sky Survey Data Release 4 \cite{Wang:2007ud}.
This suggests that halos are mostly symmetric in Cartesian projection, which is independent of the observer's distance. In the same figure, we show the distributions of axis ratio in the DM-only and the DM+baryon simulations. We only find a minor tendency for halos in the DM+baryons simulation to be more symmetric in projection.

\begin{figure*}[t] 
\begin{center}
\includegraphics[width=0.45\textwidth]{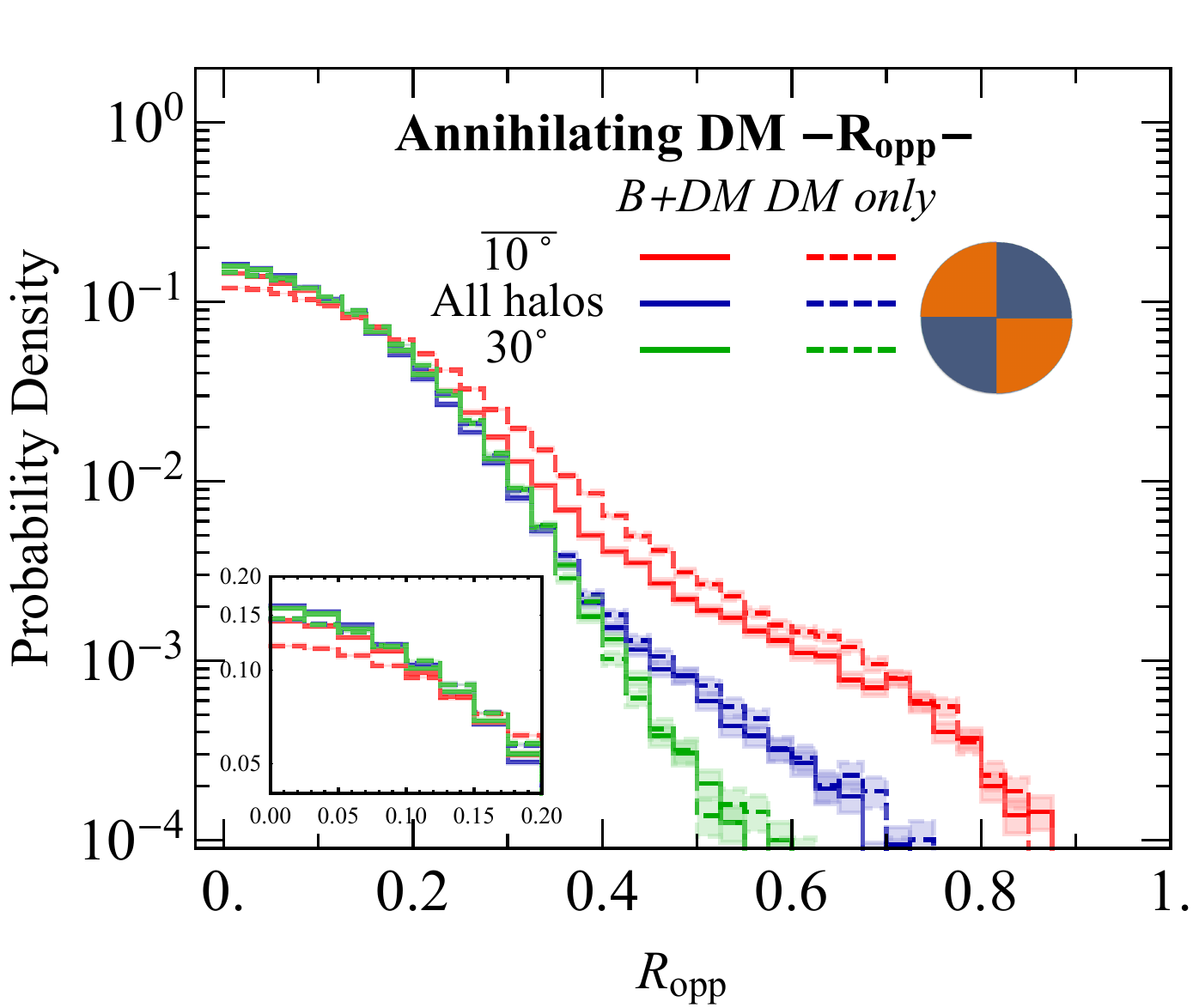}
\qquad
\includegraphics[width=0.45\textwidth]{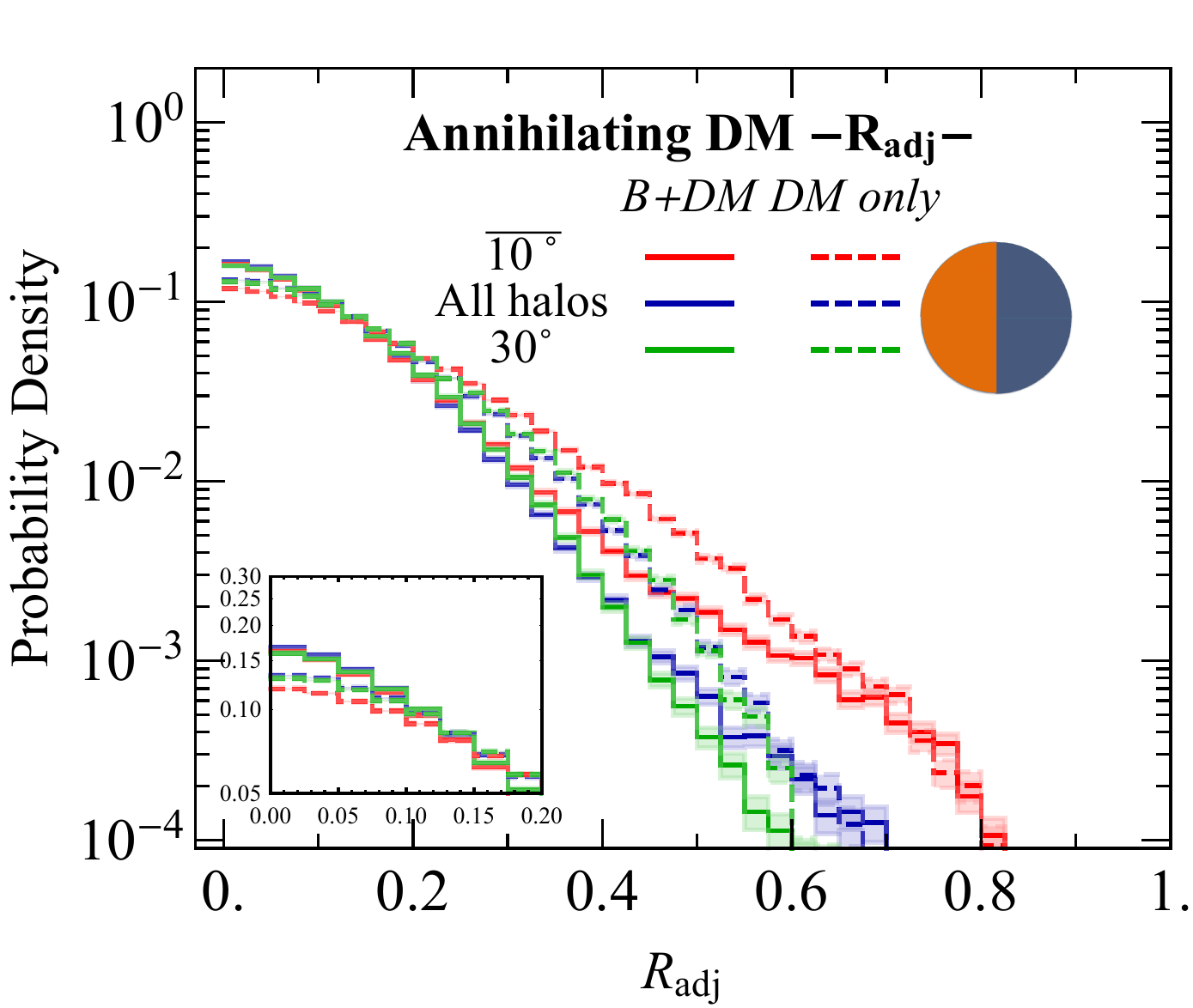}
\caption{\label{fig:illustrisann}
The distribution of the asymmetry parameters $R_\mathrm{opp}$ (left) and $R_\mathrm{adj}$ (right) for DM annihilation as observed from a point 8.5 kpc from the halo center, in halos taken from the DM-only and DM+baryon simulations of Illustris-1. $J$-factors are computed over all halos (blue) as well as when omitting the inner $10^\circ$ disk (red), and through the inner $30^\circ$ only (green). All regions are centered on the halo center. The inset shows a zoom-in of the region of small $R_\text{opp}$/$R_\text{adj}$.
}
\end{center}
\end{figure*}

Moving to the $\mathcal{J}$-tensor, we plot the distributions of the observed axis ratio for indirect detection signals in Fig. \ref{fig:axis_ratio} (right), for observers at distances from the center of the halo comparable to the solar circle radius.\footnote{Since in this figure we include all halos, not only MW-sized halos, the solar circle radius may be much smaller or larger as a fraction of the virial radius than it is in the Milky Way. However, we have checked that for all but a handful of halos, the observer is still within the halo virial radius at this distance; furthermore, we show results specifically for MW-sized halos in Sec. \ref{sec:mwhalos}.}
These distributions are generally more peaked towards higher axis ratios, with the peak now at axis ratio $\approx 0.9$. The distributions for annihilation signals are broader than those for decay; annihilation signals accentuate anisotropies due to their dependence on the square of the DM density. In the case of annihilating DM, the effect of including baryons is more pronounced; DM-only halos are less spherical/more elongated than those including baryons. This effect has been previously studied and our results are consistent with previous analyses \cite{1991ApJ...377..365K,Dubinski:1993df,Debattista:2007yz,Pedrosa:2009rw,Tissera:2009cm,Zemp:2011nk,2011ApJ...740...25B,Bryan:2012mw,Kelso:2016qqj}.

\subsection{Quadrant Analysis}
\label{sec:IGallhalos}
In this section, we compute the distributions of $R_{\text{opp}}$ and $R_{\text{adj}}$ as defined in Eqs. \ref{eq:radj} and \ref{eq:ropp}, for the halos of Illustris-1 and Illustris-1-Dark. This measure is of particular interest as it is easy to test in template analyses (See for example Refs. \cite{Finkbeiner2004,Daylan:2014rsa}). For every halo, we build a map of the $J$-factors at different points in the sky in annihilation and decay following the procedure in Sec. \ref{sec:jfactor}. We picked nside $= 16$ (see Eq.~\ref{eq:nside}), but tested that the results are stable under a change of nside. In order to characterize the spatial distribution of indirect detection signals, we consider three different regions of each halo. First, we take the halos as a whole, then we omit the inner cone of half angle $10^\circ$ and finally we look at the inner cone of half angle $30^\circ$. 10 degrees from the perspective of an observer at a distance of 8.5 kpc covers one softening length $\epsilon_\mathrm{DM} = 1.4$ kpc . 
We show the results for DM annihilation in Fig. \ref{fig:illustrisann} as a probability distribution 
of $R_{\text{opp}}$ and $R_{\text{adj}}$. Similar distributions can be found for decay as shown in App. \ref{app:decay}. We have included Gaussian error bars, shown as the shaded regions of Fig. \ref{fig:illustrisann}. 

For the case of annihilation, shown in Fig. \ref{fig:illustrisann}, we find that there is a notable difference in the sphericity of the DM-only simulation compared to the DM+baryons simulation; there are more asymmetric halos ($\Delta J/J_{\text{total}} \gtrsim 0.3$)\footnote{By $\Delta J/J_\text{total}$ we mean $R_\text{opp}$ and $R_\text{adj}$.} in the DM-only case. This result confirms the finding in the previous section, that inclusion of baryons tend to make DM halos more spherical, although in both cases, most halos have mostly spherical decay/annihilation signals. 

 In order to understand the effect of the inner 10 degrees,  we compare the histogram of the whole halo minus the inner 10 degrees with that of the distribution that includes the entire halo. We find that deviations only happen at  larger values of $R_\text{opp}$ and $R_{\text{adj}}$ which occur with probabilities less than a few percent. This leads us to believe that our distributions are largely unaffected by the inner few softening lengths where resolution artefacts might play a larger role. 
 The innermost region of the halo tends to also be more spherical than outer regions, as shown in Fig. \ref{fig:illustrisann} when comparing the distribution in which we omitted the inner $10^\circ$ with the distribution across the whole halo.
The distributions of $\Delta J /J_{\text{total}} $ are slightly more peaked towards zero in the inner cone of half angle 30 degrees. The differences between the three regions intensify in the tail of the distribution. We note that due to the resolution of the simulation, we cannot make a statement on the sphericity of the inner few degrees around the Galactic Center.

\subsection{Correlation with Baryon Disk}
\label{sec:baryons}

In this section, we examine the distribution of the angle $\theta$ between the angular momentum vector of the baryonic disk and halo's minor axis found in projection. We first outline how to compute each of these axes. 

In order to find the orientation of the  baryon disk, we compute the three dimensional angular momentum vector of the star forming gas (SG). We first determine the location of the gas particles with a positive star forming rate, since the gas must have cooled to form stars and contribute to the angular momentum of the disk, and we then compute the 3D angular momentum vector $\vec{L}_3$ for a particular halo as
\begin{equation}
\vec{L}_3 = \sum_{i \in \text{SG}} m_{\rm{gas}}\, \vec{r}_i \times \vec{v},
\end{equation}
where $\vec{r}_i =  \vec{x}_i - \vec{x_0}$, with $\vec{x}_i$ ($\vec{x_0}$) the coordinates of the particle $i$ (the center of the halo), and $m_{\rm{gas}}$ ($\vec{v}$) is the mass (3D velocity) of the gas cell. We then project the angular momentum vector $\vec{L}_3$ on the plane perpendicular to the line between the observer and the halo, and label the new 2D angular momentum vector $\vec{L}$.

We now turn to computing the halo's minor axis. As shown in Sec. \ref{sec:axis_ratio}, there are multiple ways to compute the inertia tensor in projection: (1) projecting the particle coordinates onto the plane perpendicular to the line between the observer and the center of the halo, (2) computing the $\mathcal{J}$-tensor in annihilation and (3) computing the $\mathcal{J}$-tensor in decay. In each of these cases, we compute the eigenvalues and eigenvectors of the tensor. The eigenvector that corresponds to the smallest eigenvalue  is taken to be the halo's minor axis $\vec{M}$. 

\begin{figure}[t]
\begin{center}
\includegraphics[width=0.45\textwidth]{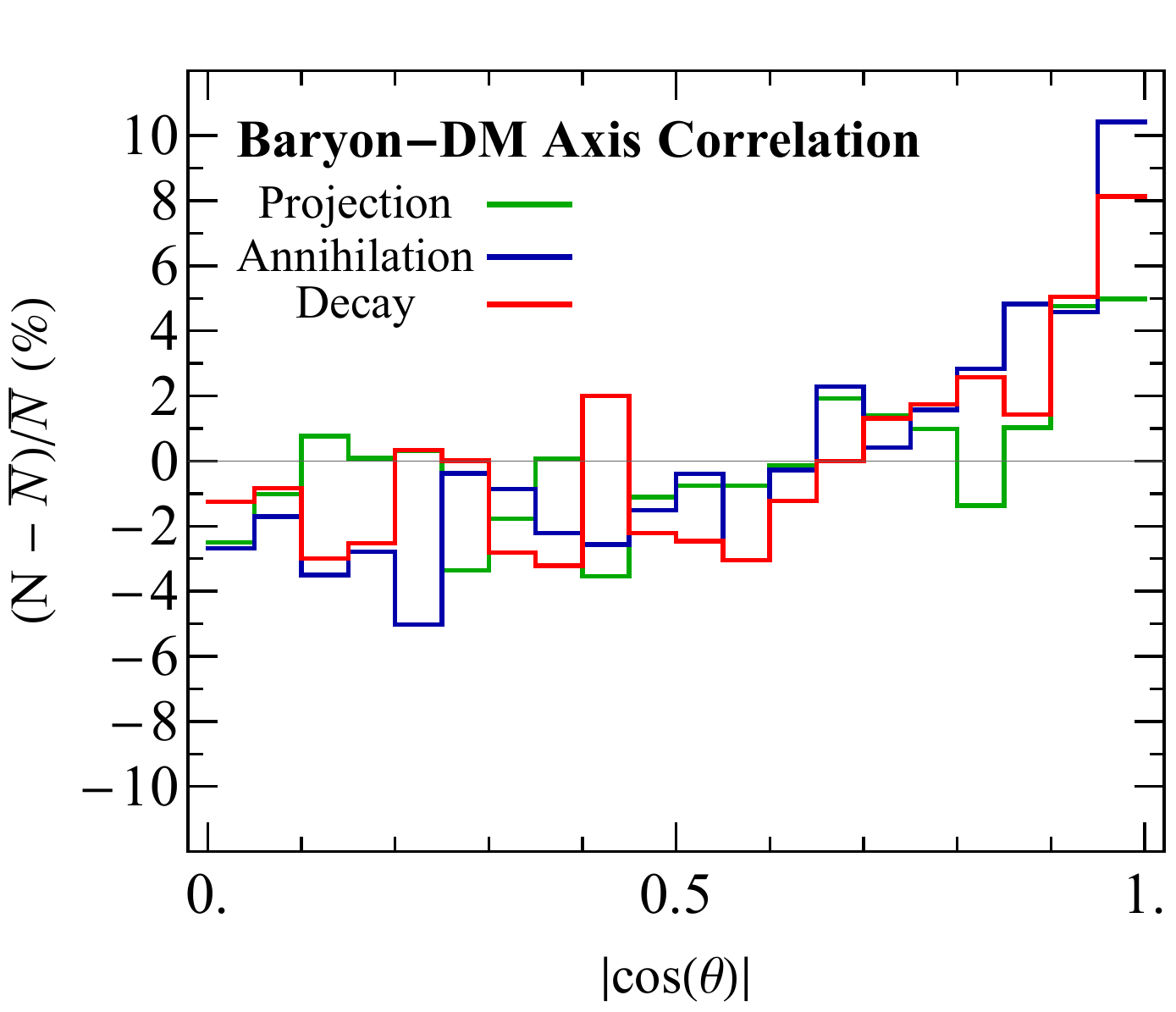}
\caption{\label{fig:anglescorrelation}
Histogram of the angle between the minor axis of the DM halo and the angular momentum vector of the star forming gas. The minor axis is the eigenvector corresponding to the smallest eigenvalue of the inertia tensor found from regular projection (Eq. \ref{eq:inertiatensor}) or the weighed inertia tensor (Eq.~\ref{eq:jfactortensor}) in the case of annihilation and decay. $\overline{N}$ is the mean of the distribution (a flat distribution in $\cos \theta$).
}
\end{center}
\end{figure}

In order to compute the angle between the minor halo axis and the angular momentum vector, we consider the normalized inner product $\vec{M} \cdot \vec{L}/(|\vec{M}| |\vec{L}|) = \cos \theta$. In Fig.~\ref{fig:anglescorrelation}, we compare the newly found distribution of the angle $\theta$ to that of a flat distribution in $\cos \theta$ using the three definitions of the minor axis. We find that these three measures are consistent with a slight correlation between the halo's minor axis (found in projection) and the angular momentum vector; there is a slight preference for the two axes to be aligned to each other, i.e. $|\cos \theta| \sim 1$. This implies a slight preference for the halo's major axis to be aligned with the baryonic disk.  

Our results found in projection are consistent with previous 3D analyses \cite{Bailin:2005xq,AragonCalvo:2006ay,Hahn:2010ma,Debattista:2013rm,Velliscig:2015ffa,2012MNRAS.420.3303B,Codis:2014awa,Dubois:2014lxa,Chisari:2015qga,Velliscig:2015ixa,Kiessling:2015sma,Debattista:2015hia,2011MNRAS.413.1973W} (Ref. \cite{Tenneti:2014poa} shows a 2D projected misalignment angle of order $10^\circ$).

\subsection{Milky-Way-like Halos}
\label{sec:mwhalos}

\begin{figure}[t]
\begin{center}
\includegraphics[width=0.45\textwidth]{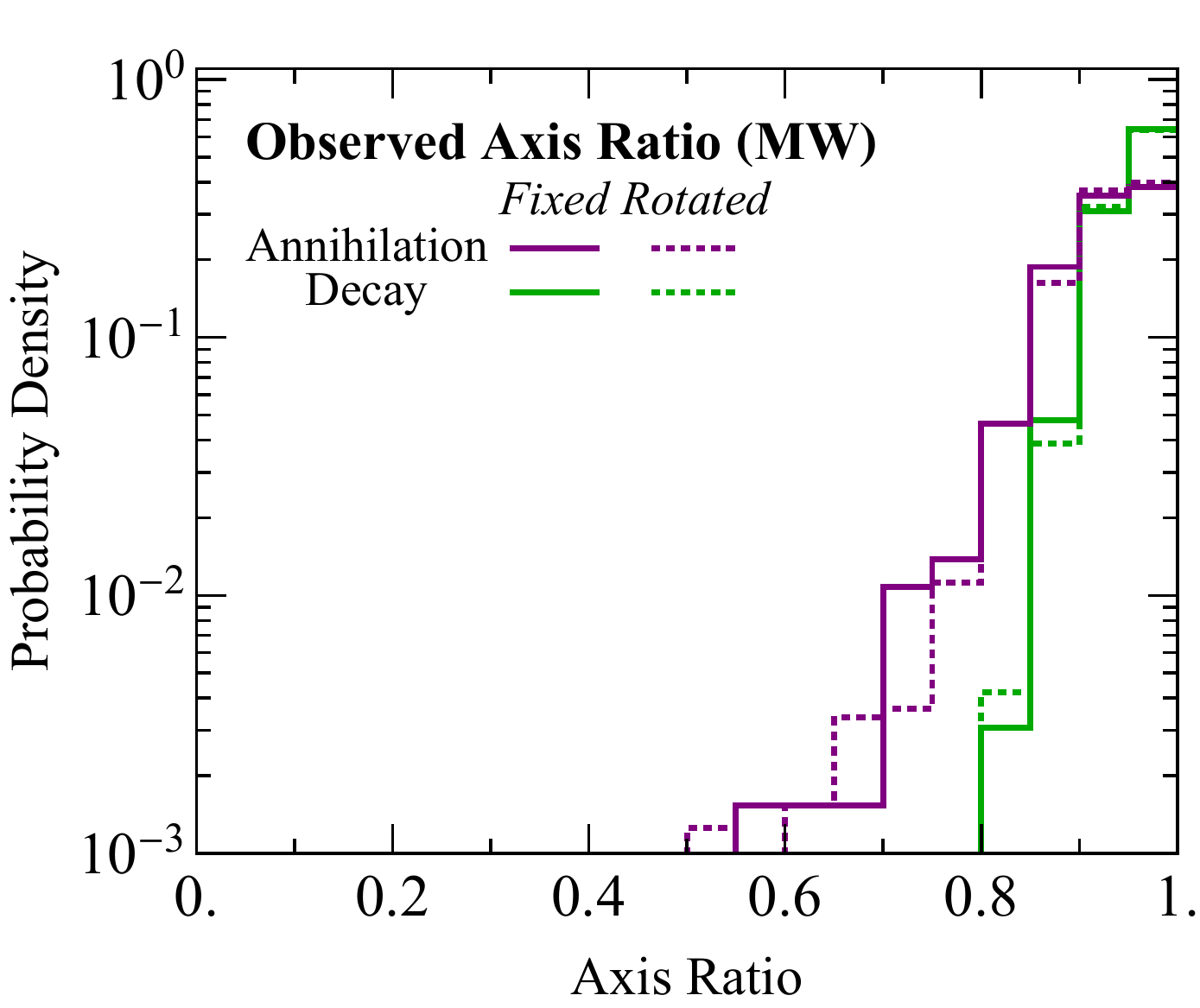}
\caption{\label{fig:axis_ratio_MW}
 Histogram of the observed axis ratio for annihilation and decay of MW-like halos as defined by the requirements in Eqs. \ref{eq:totalmass} and \ref{eq:starmass}. The distribution of the MW-like halos is shown in solid lines while the distribution of rotated halos to increase statistics in shown in dotted lines.
}
\end{center}
\end{figure}

We now focus on the subset of MW-like halos, to see if they share consistent sphericity properties with the overall sample. This is crucial, as were we to discover DM through its annihilation/decay to SM particles in the MW, the signal/background could be analyzed exactly in the same way we analyze the Illustris data. To that aim, we require the following:
\begin{itemize}
 \item Total mass: The total mass of the halo lies in the range (see for example Ref. \cite{Schaller:2015mua})
 \be \label{eq:totalmass}
 5 \times 10^{11} M_{\odot} < M_{200} < 2.5 \times 10^{12} M_{\odot},
 \ee
 where $M_{200}$ is the mass of the halo enclosed in a sphere with a mean density 200 times the critical density of the Universe today. $M_{\odot}$ is the solar mass.
The number of halos in Illustris-1 within this mass range is 1652.
 \item Stellar mass: The total stellar mass lies within the range \cite{McMillan:2011wd,Calore:2015oya}
 \be \label{eq:starmass}
 4.5 \times 10^{10}  M_{\odot} < M_{\text{Stars}} < 8.3 \times 10^{10}  M_{\odot}.
 \ee
This further drops the number of MW-like halos in the Illustris-1 simulation to 650.
\end{itemize}
We then perform the analysis of Sec. \ref{sec:obs_axis_ratio} on this restricted sample of halos. We find that indeed the distributions shown in Fig. \ref{fig:axis_ratio_MW} are consistent with the more general results shown in Fig. \ref{fig:axis_ratio}, though with lower statistics. The DM signal is expected to be spherical, and peaks at values $\approx 0.8 -0.9$, although with a more peaked distribution. 

 In order to increase statistics, we study the observed axis ratio for MW-like halos from 12 different projections by placing the observer at different locations along the sphere centered at the halo center with radius $R_\odot$~kpc. We plot the new distribution in Fig. \ref{fig:axis_ratio_MW} in dotted lines. We find that the distributions are preserved but more smoothed out.

We will discuss in Sec. \ref{sec:fermi} the axis ratio of the gamma-ray sky as observed by \textit{Fermi} \cite{Atwood:2009ez}, as compared to the distribution of observed axis ratios in MW-like DM halos.

\section{Extragalactic Analysis}
\label{sec:extragalactic}

In this section, we perform a similar analysis to Sec.~\ref{sec:results}, but we now situate the observer outside the halo in consideration. As an example, we set the observer at a distance
\begin{equation} \label{eq:distance}
 r = 2~R_{200},
\end{equation}
where $R_{200}$ is the distance from the halo center at which the overdensity of the halo is 200 times the critical density of the Universe. We check that our results are independent of the distance between the observer and the center of the halo as long as $r > R_{200}$. We increase nside to 512 in this analysis to be able to resolve smaller structures of the halos (See Eq. \ref{eq:nside}), then downgrade the maps to nside = 32 for computational efficiency in the analysis.\footnote{If the maps are generated originally at nside=32, the lines of sight through the center of each pixel do not adequately describe the average emission from that pixel, as large variations in the brightness can occur on scales smaller than a pixel. Consequently, changes in the pixelation can markedly change the results. To resolve this problem, we generate the maps at higher resolution, and use these higher-resolution maps to determine the total emission in each (nside=32) pixel. Once this is done, our results are stable with respect to the choice of pixel size.} With this choice of $r$, the halos cover $\sim 30^\circ$ of the map, which is higher than most extragalactic signals, but we do so in order to resolve the inner structure. 

\begin{figure}[t]
\begin{center}
\includegraphics[width=0.45\textwidth]{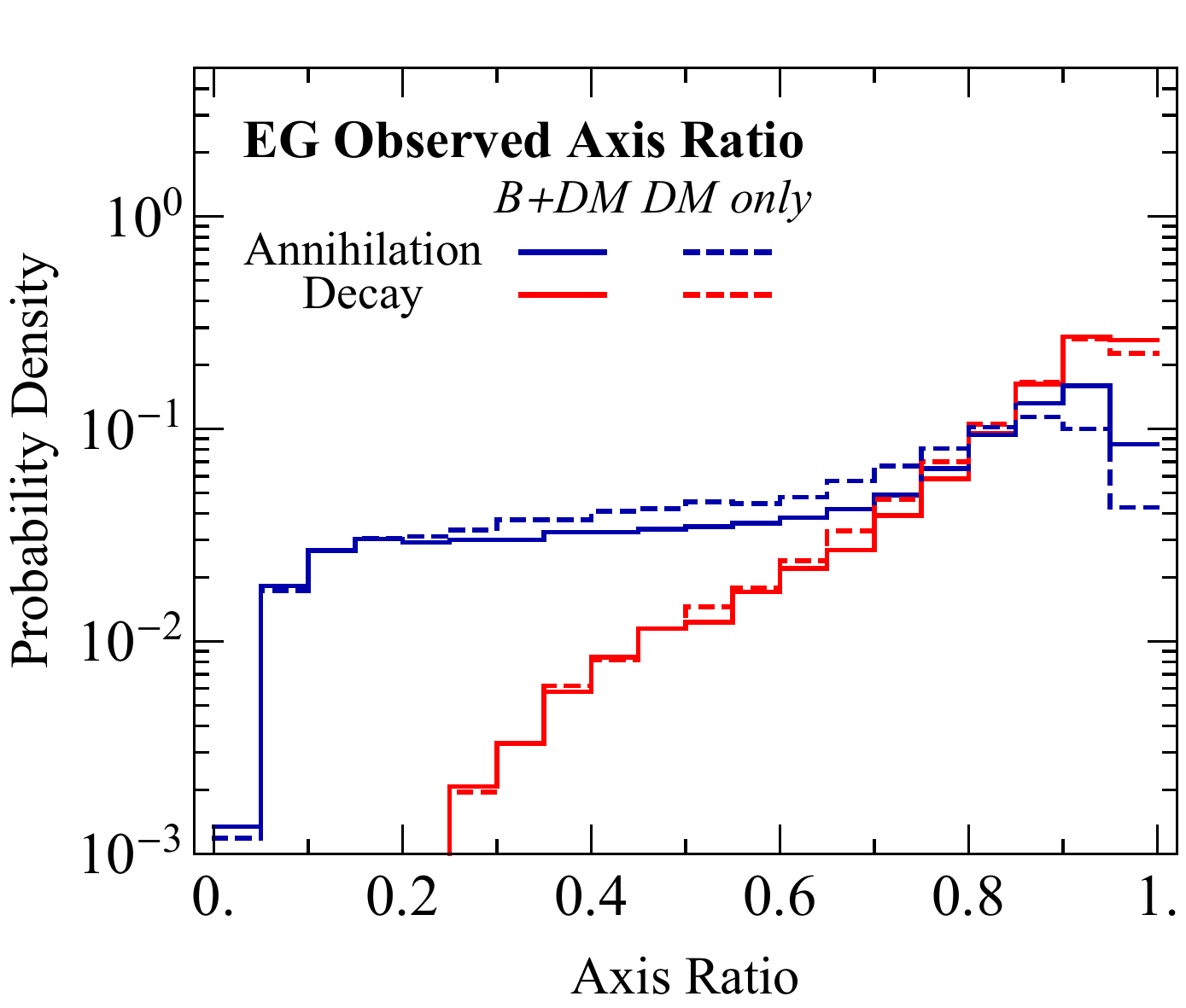}
\caption{\label{fig:axis_ratio_EG}
Histogram of the 
observed axis ratio for annihilation and decay for extragalactic sources, comparing both Illustris-1 and Illustris-1-Dark (see Eq. \ref{eq:jfactortensor}).}
\end{center}
\end{figure}

\begin{figure*}[t] 
\begin{center}
\includegraphics[width=0.45\textwidth]{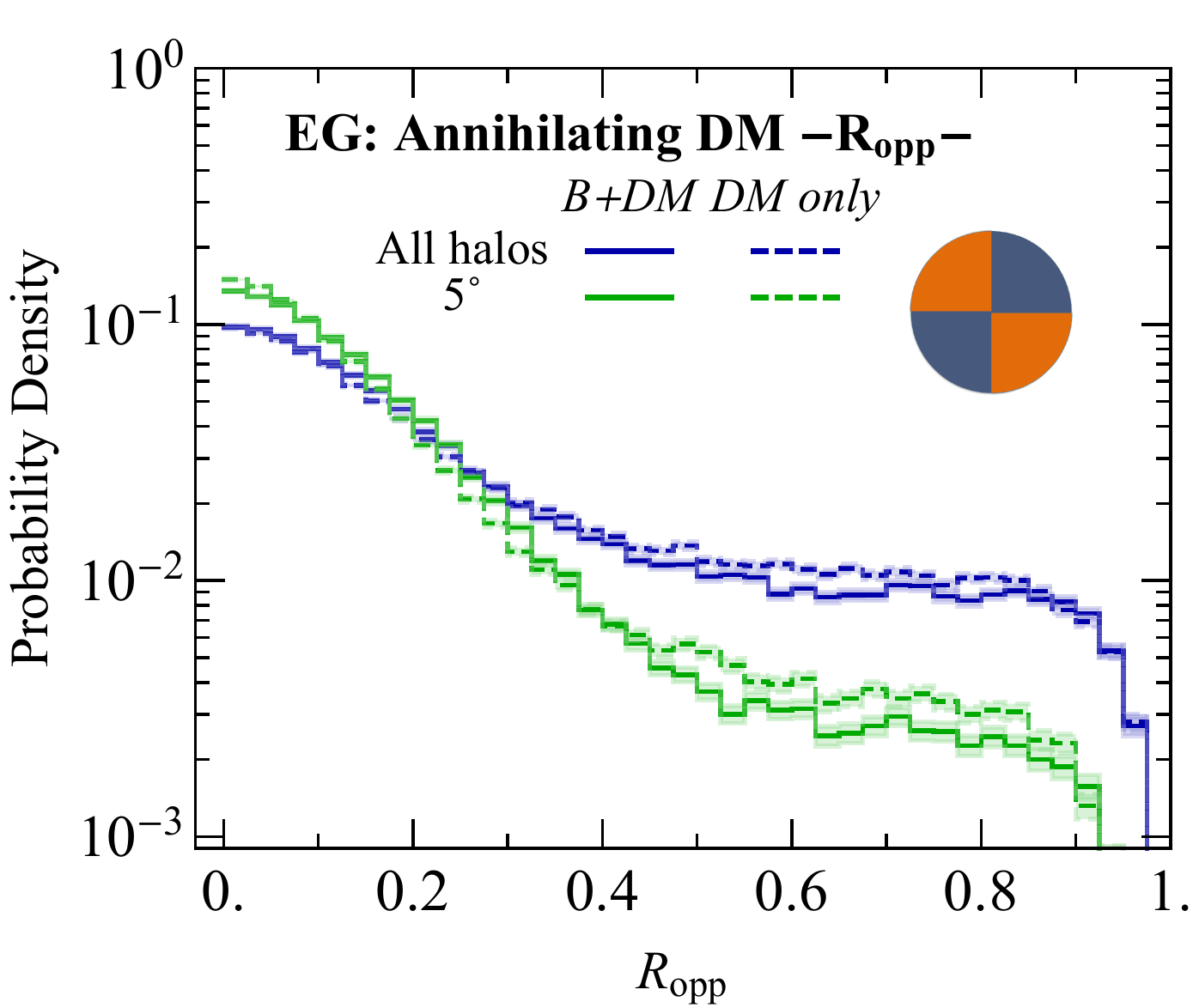}
\qquad
\includegraphics[width=0.45\textwidth]{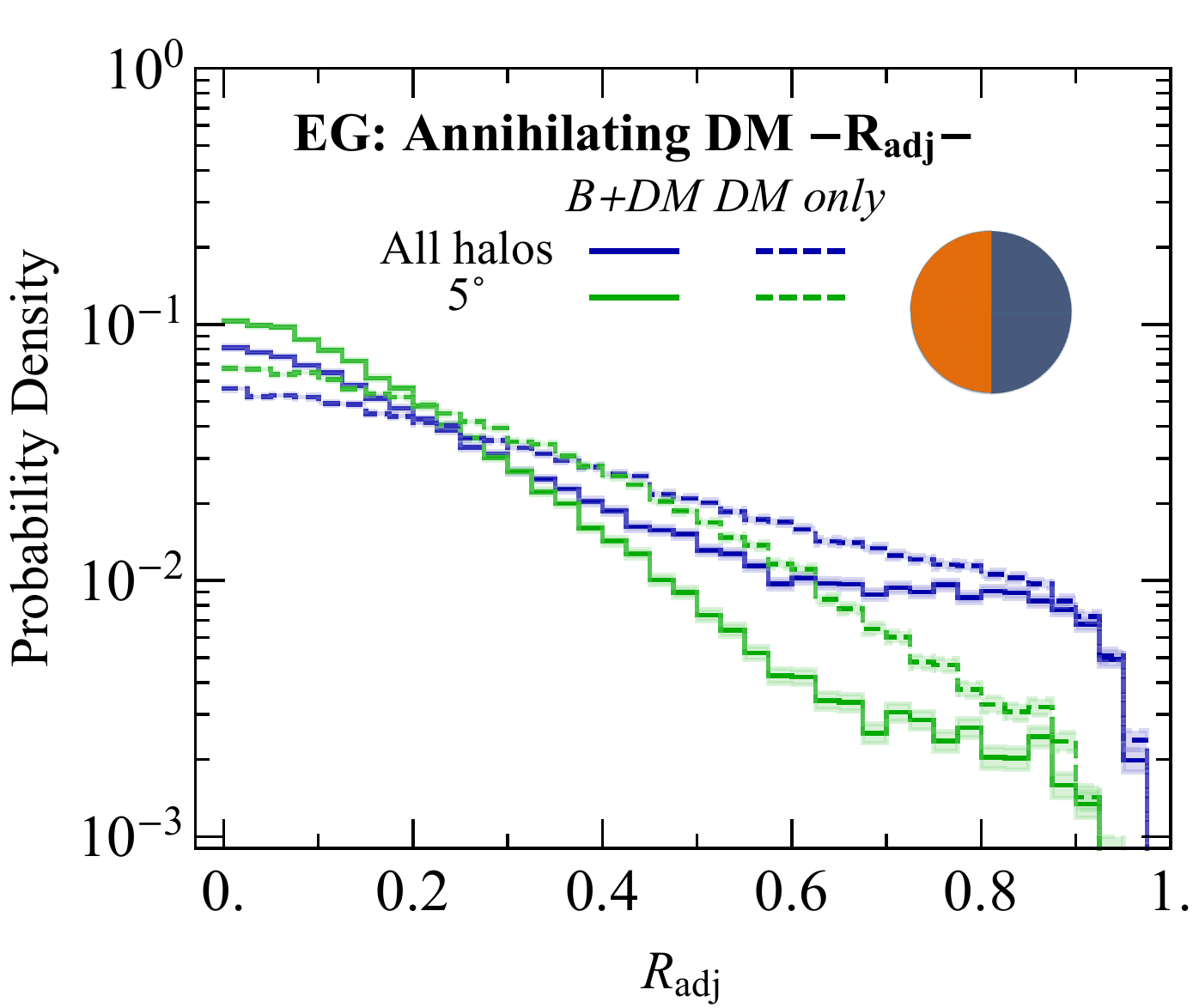}
\caption{\label{fig:illustrisEGann}
The distribution of the asymmetry parameters $R_\mathrm{opp}$ (left) and $R_\mathrm{adj}$ (right) for DM annihilation as observed from a point outside the halo, taken from DM-only and DM+baryons simulations. $J$-factors are computed over all halos (blue) and through the inner $5^\circ$ only (green).
}
\end{center}
\end{figure*}

\subsection{Axis Ratio}
\label{sec:axis_ratio_EG}

As with our previous analysis (Sec. \ref{sec:results}), we study the distribution of both the relevant axis ratios and the quadrant parameters. In Fig. \ref{fig:axis_ratio_EG}, we plot the distribution of observed axis ratio in the case of annihilation and decay of DM, for the DM-only simulation as well as DM+baryons simulation. In both decay and annihilation, the distributions of axis ratio are flatter than in the Galactic analysis; while decay signals still generally have fairly spherical profiles, the distribution of axis ratio for annihilation signals is nearly flat, although slightly peaked around 0.9. An interesting feature is the non-negligible fraction of halos with axis ratio $0.1 - 0.4$. As we will explore in Sec. \ref{sec:mergers}, this behavior is due to halo mergers.

Serving both as a consistency check and as a study of the baryonic effects, the DM-only simulation exhibits similar features to the baryonic simulation, with the distributions shifted slightly towards lower values of the axis ratio. 

\subsection{Quadrant Analysis}
\label{sec:quad_EG}

As shown in Fig. \ref{fig:illustrisEGann}, the ratios of opposite and adjacent quadrants show that DM signals are less spherical when observed at a larger distance. This is reasonable as all features of the halo are at an equivalent distance from the observer, while in the Galactic analysis, it is harder to resolve small anisotropies that are at a larger distance from the observer. These results suggest that especially for extragalactic annihilation signals, observation of an elongated morphology could not be used to disfavor a DM hypothesis, and there is no reason to expect highly spherical signals that could easily be distinguished from astrophysical sources with complex and non-spherical distributions. (However, if the primary astrophysical backgrounds were near-spherical, a highly elongated profile might provide a hint for a DM origin.)

In order to omit possible signals from secondary subhalos which are off the center of the halo, defined by the most bound particle, we analyze the ratios of the quadrants within a cone of half angle $5^\circ$. We find that the distributions within the cone do indeed appear more spherical, but the effect generally dominates at the tail of the distribution, where the asphericity is more extreme.

\begin{figure}[t]
\begin{center}
\includegraphics[width=0.45\textwidth]{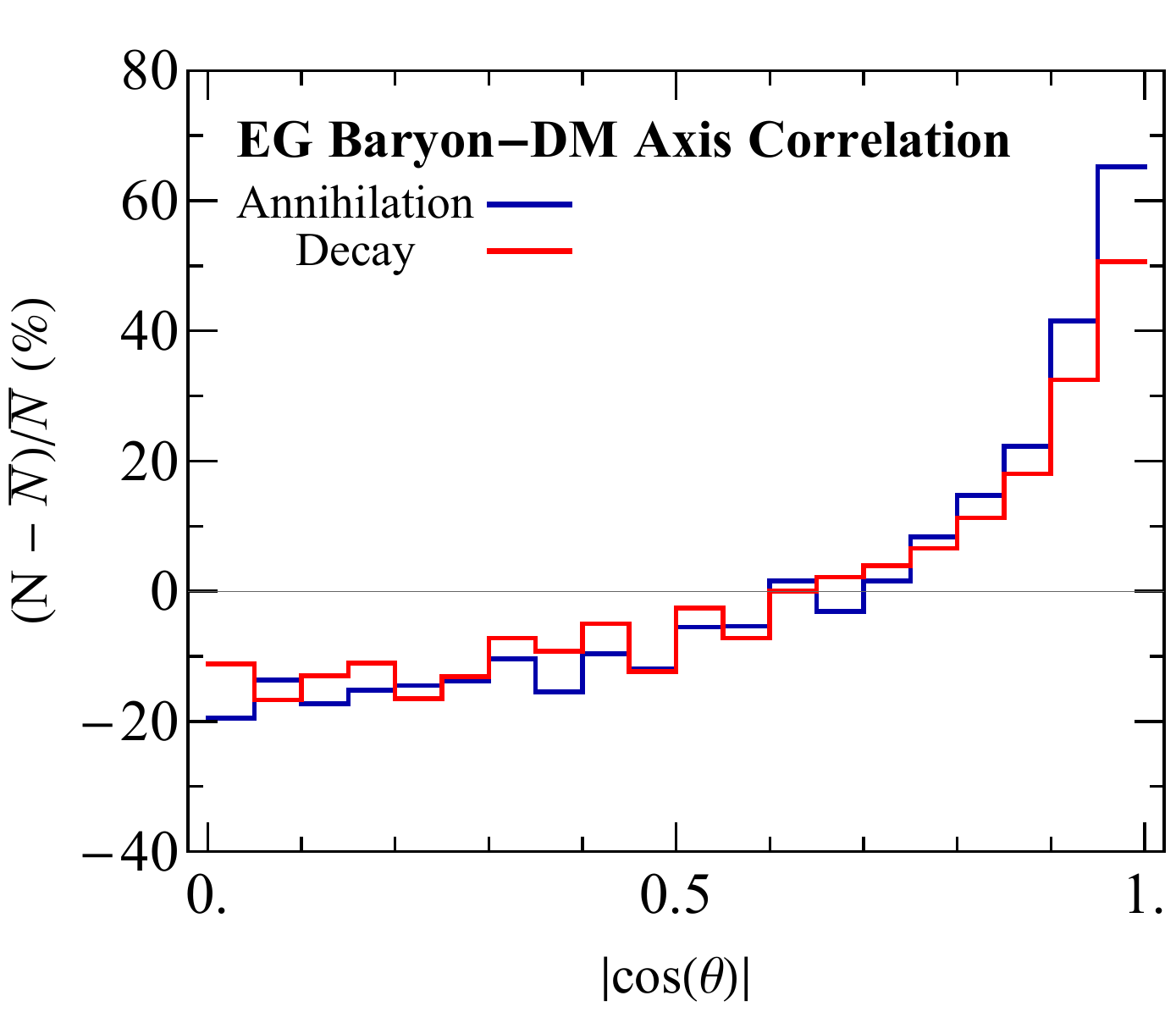}
\caption{\label{fig:anglescorrelation_EG}
Histogram of the angle between the minor axis of the DM halo for the case of extragalactic signals and the angular momentum vector of the star forming gas. The minor axis is the eigenvector corresponding to the lowest eigenvalue of the inertia tensor found from regular projection (Eq. \ref{eq:inertiatensor}) or the weighed inertia tensor (Eq. \ref{eq:jfactortensor}) in the case of annihilation and decay. $\overline{N}$ is the mean of the distribution (a flat distribution in $\cos \theta$).}
\end{center}
\end{figure}

\subsection{Correlation with Baryon Disk}
\label{sec:anglecorrelation}
Similarly to Sec. \ref{sec:baryons}, we plot in Fig. \ref{fig:anglescorrelation_EG} the distribution of the angle between the halo's minor axis (found in annihilation and decay of the DM particles) with the angular momentum vector of the star forming gas, this time analyzing the minor axis from the extragalactic maps. We find a strong correlation between the DM minor axis and the angular momentum vector, as the two tend to be aligned. Therefore, the halo's major axis is tangent to the baryonic disk. This is more obvious in this analysis compared to the Galactic analysis of Sec. \ref{sec:baryons} since Galactic DM signals are more spherical and therefore harder to orient in a particular direction; correlating the direction of a mostly spherical signal is done at random (See Sec. \ref{sec:baryons} for a comparison with previous work.).

\subsection{Halo Mergers / Subhalos}
\label{sec:mergers}

\begin{figure}[t]
\begin{center}
\includegraphics[trim={0 0 0 1.6cm},clip,width=0.6\textwidth]{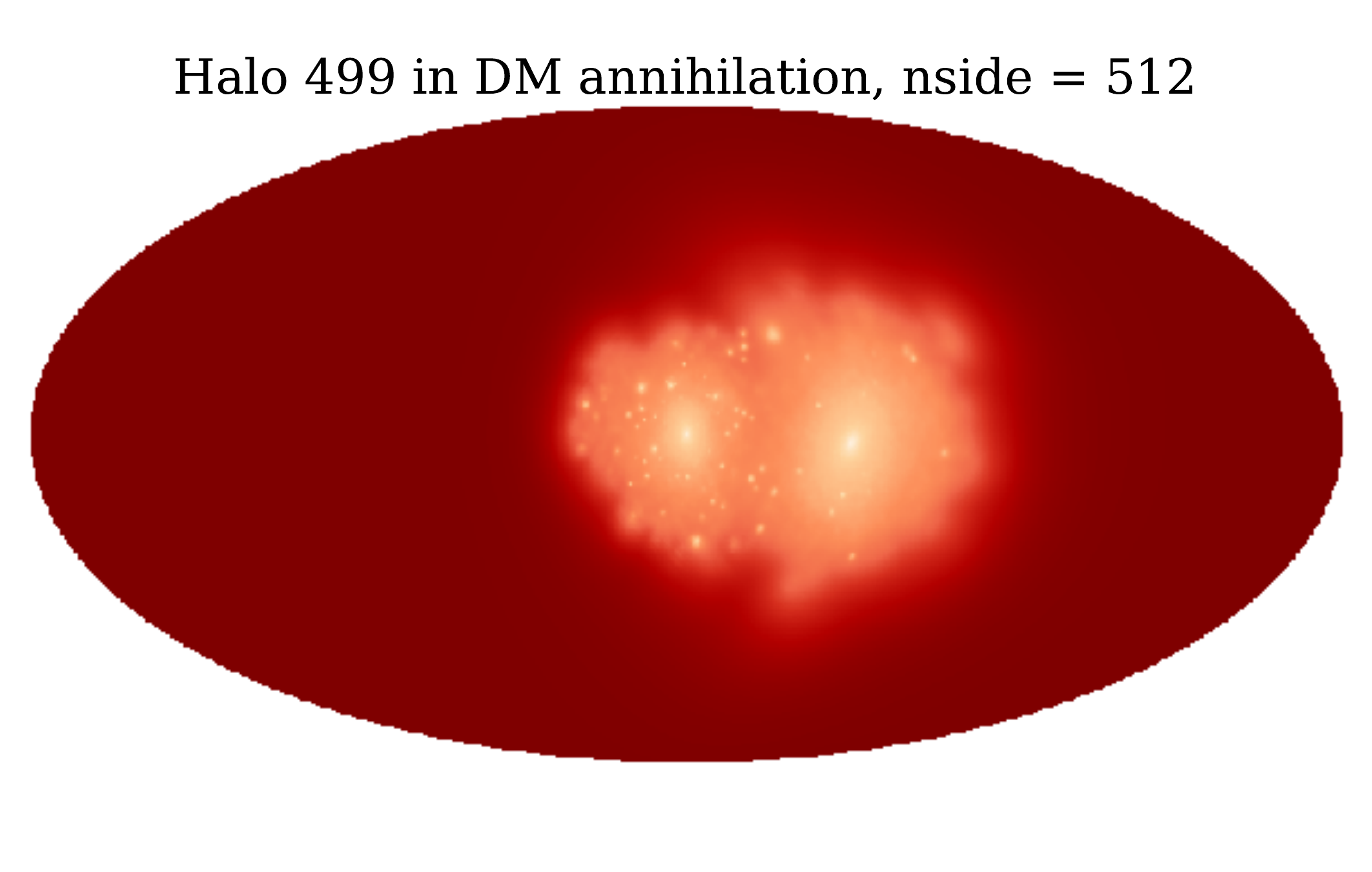}
\vspace{-10 mm}
\caption{\label{fig:merger} Logarithmic map for DM annihilation for halo labeled ``499'' in the Illustris-1 simulation from the point of view of an  observer external to the halo. We used HEALPix with nside $= 512$ (see Eq. \ref{eq:nside}). The observer is located at a distance $2 R_{200} = 478 $ kpc of the center. The halo mass is $2.72 \times 10^{12} M_{\astrosun}$. 
}
\end{center}
\end{figure}

Many of the halos in the simulation have experienced a recent merger or interaction; an example is shown in Fig.~\ref{fig:merger}. To test the effect of these mergers/large subhalos on our sphericity distributions, we study the observed axis ratio in two different sets of subsamples of the data.

\begin{figure*}[t]
\begin{center}
\includegraphics[width=0.45\textwidth]{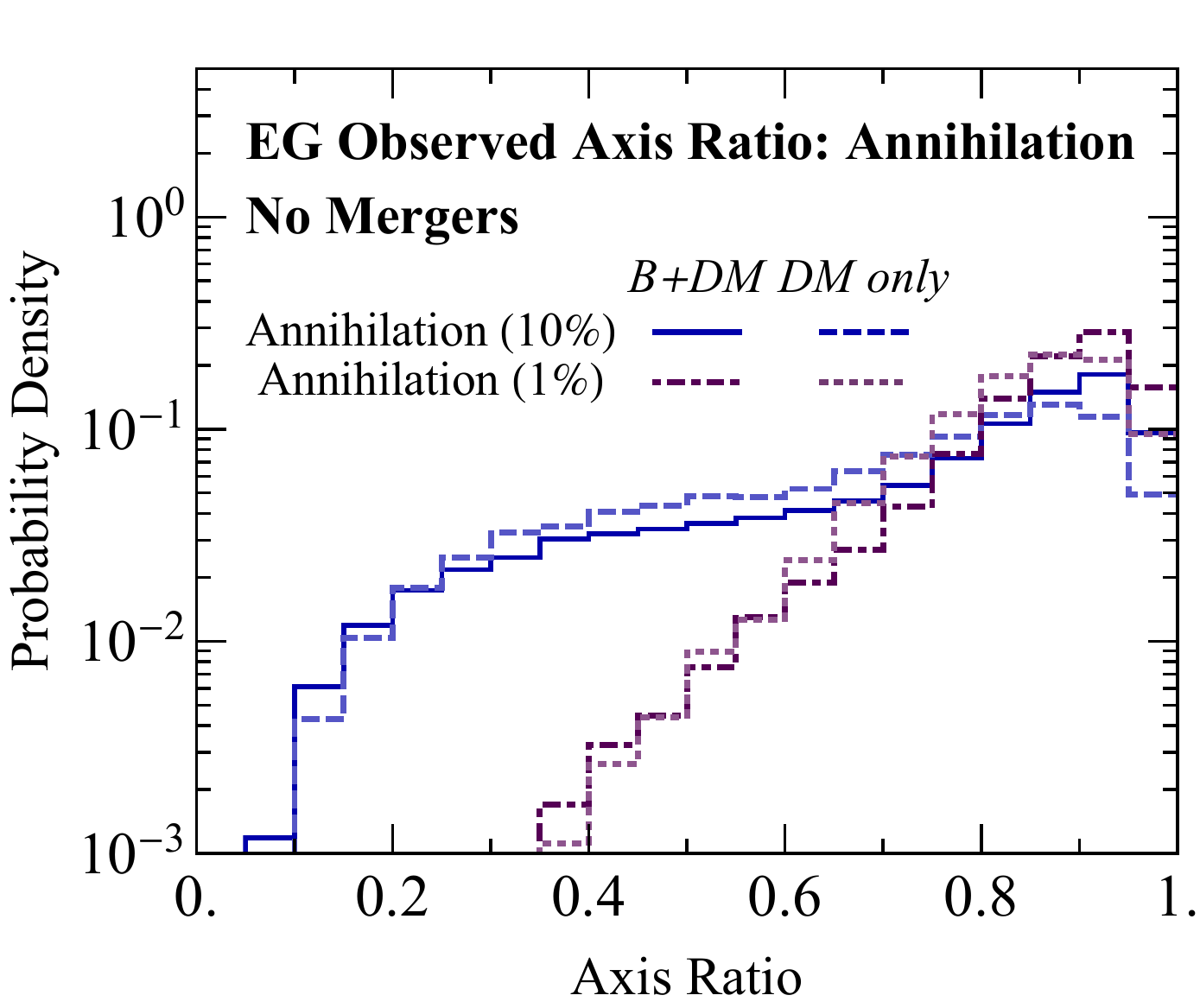}
\includegraphics[width=0.45\textwidth]{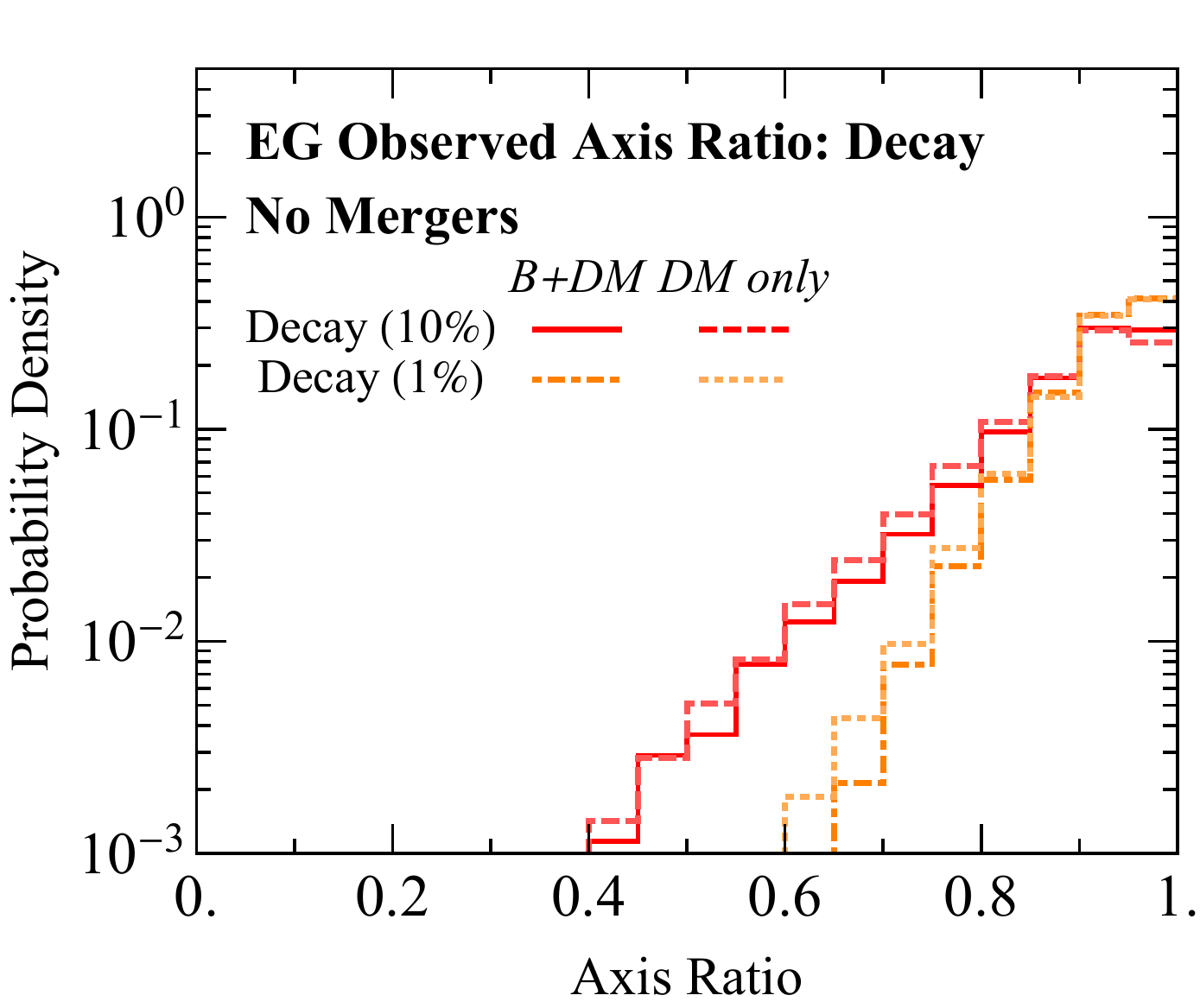}
\caption{\label{fig:axis_ratio_EG_nomerger}
Histogram of the newly defined observed axis ratio for annihilation and decay for extragalactic sources, comparing both Illustris-1 and Illustris-1-Dark (see Eq. \ref{eq:jfactortensor}) after having omitted the ``merger'' halos (see text).
}
\end{center}
\end{figure*}

First, we omit from the analysis halos where the second-largest subhalo (the first one being the host halo) has a mass fraction higher than $10\%$ ($1\%$) of the total mass of the halo. As an example, halo ``499'', shown in Fig. \ref{fig:merger}, encompasses the main host halo of mass fraction 0.49, and a second subhalo of mass fraction 0.44. Removing these halos leads to a more steeply falling axis ratio distribution for small axis ratios, $\sim0.1-0.5$, as shown in Fig. \ref{fig:axis_ratio_EG_nomerger}; compared with the original distribution in Fig. \ref{fig:axis_ratio_EG}, the low-axis-ratio longer tail of the distribution is diminished. When the cut is strengthened to remove all subhalos with more than $1\%$ of the total mass of the halo, this tail is removed almost completely.

Second, we perform the quadrant analysis on the inner $5^\circ$ of the halo, shown in Fig. \ref{fig:illustrisEGann}, which should only pick out the subhalo with the deepest potential well, as the location of halos in the Illustris simulation is set by the most bound particle. The distributions of $R_\text{opp}$ and $R_\text{adj}$ are peaked closer to zero (sphericity) when considering only pixels within the inner $5^\circ$ of the halo. The distribution is still fairly flat and not especially peaked at near-sphericity.

We see that a non-negligible fraction of the halos are expected to have elongated DM distributions due to recent mergers and/or massive subhalos. In many cases, the presence of such mergers should be apparent from the baryonic matter, but in cases where the merging halo was a low-mass system, the peak of the annihilation/decay signal might be substantially displaced from the center of the potential well inferred from the baryonic matter. This is consistent with previous work (see for example Ref. \cite{Moore:2003eq}). Alternatively, one can also try to understand the virialization of the halos through a virialization parameter such as the one given in Ref. \cite{Wise:2007jc}, though we do not do so in this work as it is computationally intensive. 

\section{Comparison to Photon Data}
\label{sec:observations}

In this section, we compare the anisotropy/sphericity distributions for DM halos, from the Illustris simulation, with the astrophysical backgrounds for potential DM signals. 

\subsection{\textit{Fermi} data}
\label{sec:fermi}
For this analysis, we use Pass 8 data from \textit{Fermi} collected between August 4, 2008 and June 3, 2015~\cite{Atwood:2009ez,Atwood:2013rka}. We employ the recommended data quality cuts: zenith angle $<90^{\circ}$, instrumental rocking angle $<52^{\circ}$, \texttt{DATA\_QUAL} $>$ 0, \texttt{LAT\_CONFIG}=1. We use the Ultraclean event class and select the top quartile of events by point spread function \cite{Portillo:2014ena}. We divide these photons into thirty equally logarithmically-spaced energy bins between 0.3 and 300~GeV; we restrict our analysis to the eight energy bins covering the range from $\sim 2-12$ GeV, as in this energy range the point spread function of the telescope is small and stable, and the gamma-ray excess has been clearly detected \cite{Goodenough:2009gk,Abazajian:2012pn,Daylan:2014rsa,Calore:2014xka,TheFermi-LAT:2015kwa}.

First, we analyze the full \textit{Fermi} data with no additional cuts, as it is dominated by background. We pixelize the sky using HEALPix with nside $=128$, and adopt the same strategy outlined in Sec. \ref{sec:results} for the analysis of signals from within a halo, where the center of the halo is located at 8.5 kpc from the observer. Computing the $\mathcal{J}$-tensor defined in Eq. \ref{eq:jfactortensor}, we find an average axis ratio of $0.54$, with a few percent spread across the different energy bins. 
In our default orientation, none of the 650 MW-like halos in the sample, shown in Fig. \ref{fig:axis_ratio_MW}, had an axis ratio this small or smaller, in either annihilation or decay. When we tested the effect of viewing the halos from different directions, we still found no halos with this level of elongation in decay signals, but for annihilation, two halos (out of 650) attained this level of elongation for specific orientations, corresponding to 10 samples out of $650 \times 12 = 7800$ tests.

Second, we isolate the residual \textit{Fermi} signal in the energy bin that dominates the signal $1.89 - 2.38$ GeV, and study its morphology. The residual signal map\footnote{We thank Nicholas Rodd for providing us with the residual maps.} is obtained through a similar analysis strategy as that used in Ref. \cite{Linden:2016rcf}. The region of interest in this analysis is $1^\circ < |b|< 15^\circ$ and $|l|< 15^\circ$, as the diffuse background templates are optimized to this region. We utilize standard template fitting methods (as in \cite{Linden:2016rcf} for example) to determine the contribution of the following templates: a uniform isotropic template, a diffuse background model by \textit{Fermi}'s diffuse model \texttt{p6v11}, a bubbles template map and an NFW template for the DM contribution. The residual map is a HEALPix map with nside=256, obtained after subtraction of the non-DM contributions with a coefficient of their best fit. We find that the  axis ratio in this region is $0.99$, confirming previous results \cite{Daylan:2014rsa,Calore:2014xka} that the signal is indeed spherical. 

\begin{figure}[t]
\begin{center}
\includegraphics[width=0.45\textwidth]{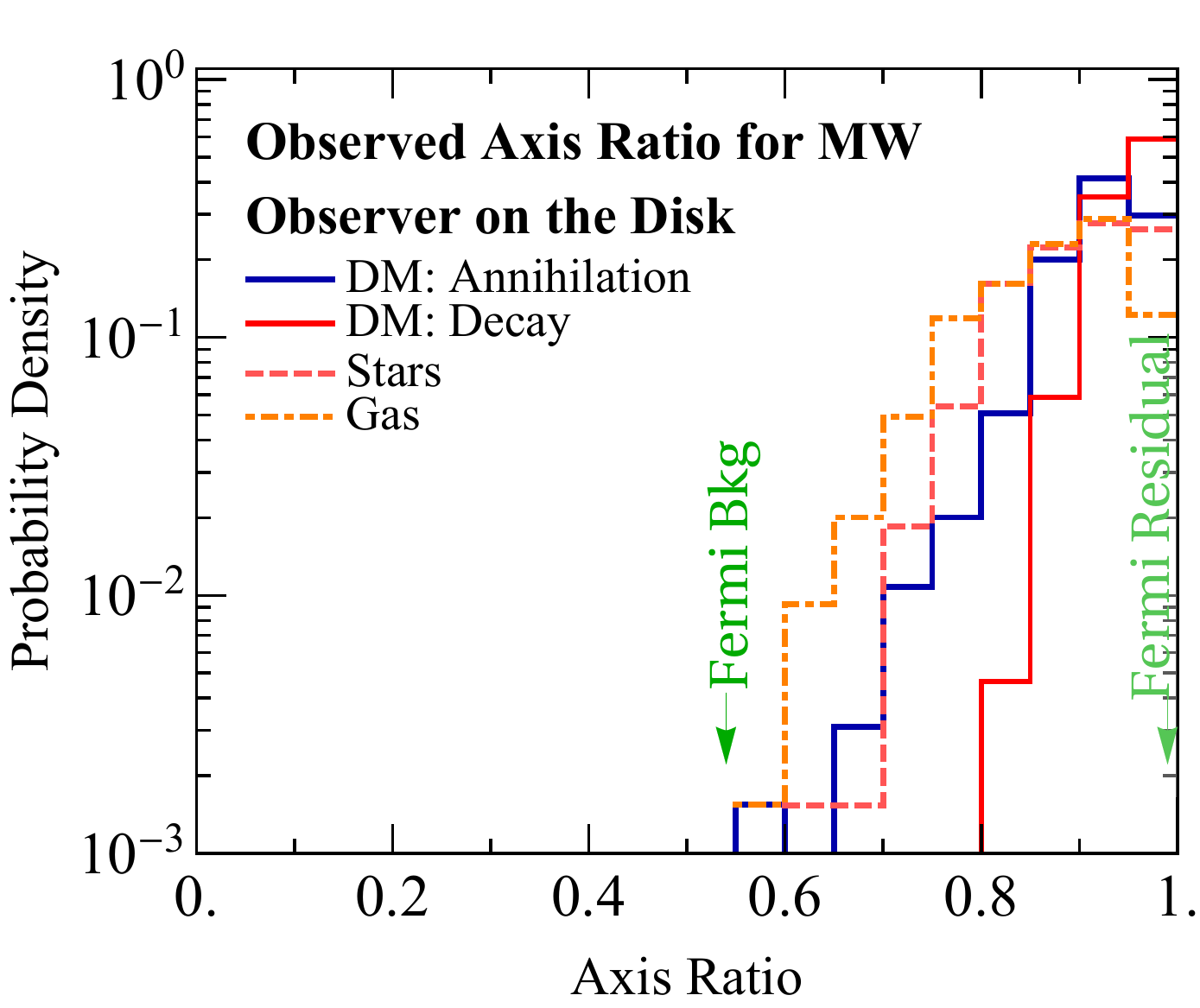}
\caption{\label{fig:axis_ratio_MW_data}
 Histogram of the observed axis ratio for annihilation and decay of DM in MW-like halos as defined by the requirements in Eqs. \ref{eq:totalmass} and \ref{eq:starmass}. We also show the histograms of the distribution of gas and stars, computed in similar manner as DM decay.  We finally show the axis ratio of the \textit{Fermi} background data, as well as the residual \textit{Fermi} signal as discussed in Sec.~\ref{sec:fermi}.
}
\end{center}
\end{figure}

For a proper understanding of the origin of the \textit{Fermi} signal and background, we perform the same analysis of Sec. \ref{sec:obs_axis_ratio} but with the distributions of gas and stars of the simulation instead of DM. \footnote{More precisely, we compute these distributions using the formalism of DM decay.} We also place the observer on the baryonic disk, defined by the plane that passes through the center of the halo and perpendicular to the angular momentum vector found in Sec. \ref{sec:baryons}. We show the histograms of the axis ratio of the star and gas in Fig. \ref{fig:axis_ratio_MW_data}. We find consistent results in which the DM is more spherical/less elongated that the gas and the stars. 
We note that the \textit{Fermi} gamma-ray emission, which largely traces the gas distribution of the Milky Way, is still quite non-spherical compared to the gas distribution of most Illustris halos. It would be interesting to understand if this reflects a general tendency for the baryonic component of Illustris halos to be more spherical and less disk-like than in reality, at least for spiral galaxies (which are known to be difficult to reproduce in cosmological simulations \cite{Vogelsberger:2014kha}). To do so one could refine the criteria imposed to select the Milky-Way-like halos defined in Sec. 3.4 (total mass and stellar mass) and even impose further constraints, e.g. the local dark matter surface density or the rotation curves. However that would have decreased even more the number of halos, limiting the present statistical analysis. Furthermore, any disk could appear ellipsoidal if observed at an angle, and it is worth noting that the angular momentum vectors computed in the analysis of Illustris have significant errors, and therefore the observer could be placed slightly off the disk.

\subsection{Cluster Data}
\label{sec:cluster}

\begin{figure}[t]
\begin{center}
\includegraphics[width=0.45\textwidth]{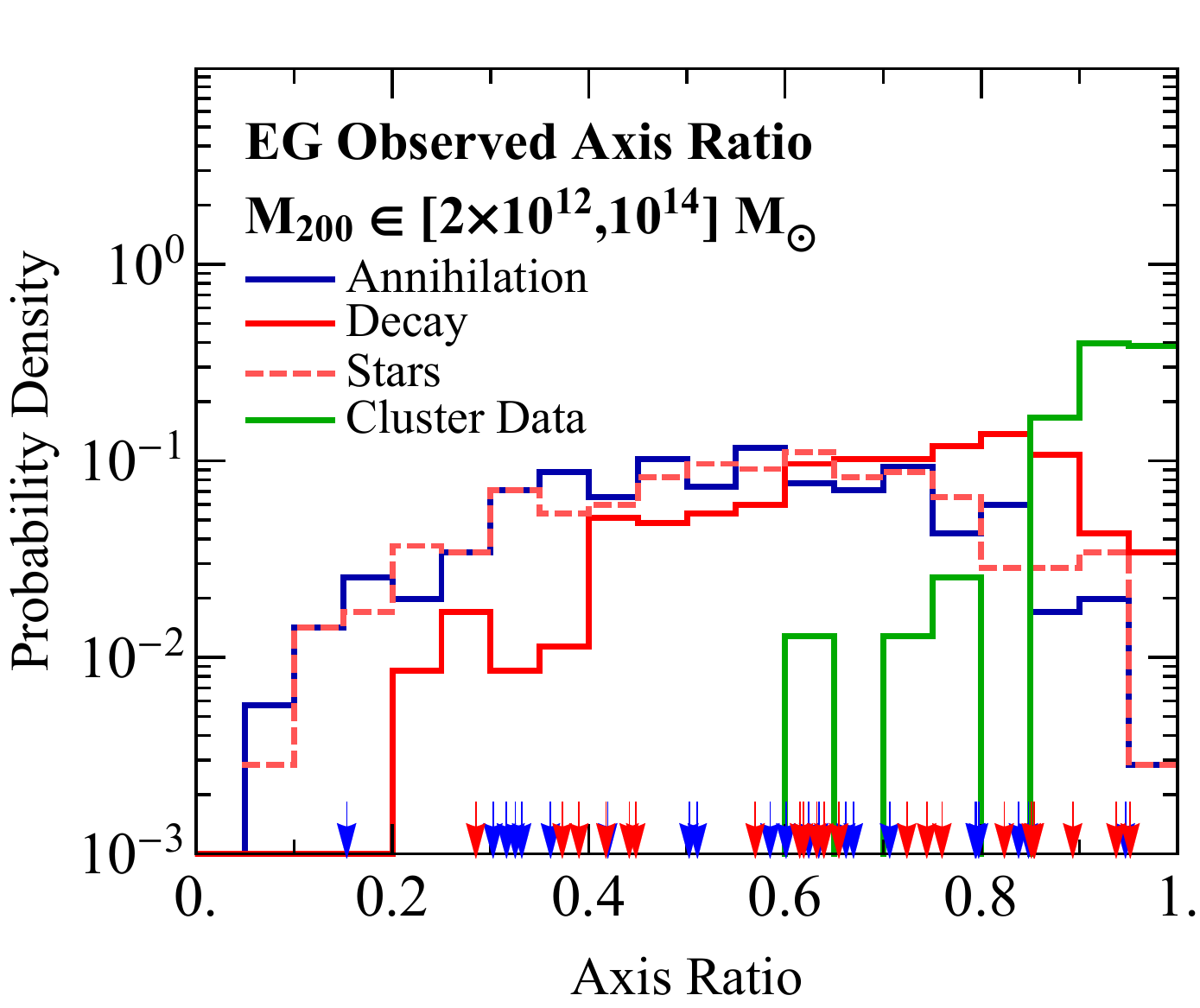}
\caption{\label{fig:axis_ratio_EGcluster}
 Histogram of the observed axis ratio for annihilation and decay of DM in cluster-like halos with masses larger than $2 \times 10^{12} M_\odot$. We also show the histograms of the distribution of stars, computed in similar manner as DM decay signals.  We finally show the axis ratio of the X-ray data. In order to make a fair comparison, we show in blue (red) arrows the location of the annihilation (decay) axis ratio of the clusters that match in mass those observed, with a cut of $M > 5 \times 10^{13} M_{\astrosun}$. We find 22 our of 352 halos that satisfy that criteria. 
}
\end{center}
\end{figure}

As an example of a potential extragalactic DM signal, we use X-ray images of 78 clusters taken by the telescope XMM-Newton \cite{Struder:2001bh,Turner:2000jy}. In a recent analysis \cite{Bulbul:2014sua,Bulbul:2014ala}, the stacked spectrum of 73 of these 78 galaxy clusters has shown a line at $E = 3.55$ keV. This sample includes clusters with a high number of counts, but to avoid the closer clusters from dominating the stacked analysis, the sample includes clusters with at least $10^5$ counts if the redshift $z<0.1$ and $10^4$ counts if the redshift is $0.1 < z < 0.4$. This sample finally yielded clusters with low redshift (less than 0.35) and masses larger than $5 \times 10^{13} M_\odot$ and therefore we compare them to Illustris maps computed at $z=0$. 

The X-ray images are obtained from XMM-Newton data\footnote{We thank Esra Bulbul for providing us with the X-ray images. She should be contacted for any image requests.}, with a field-of-view with radius 14 arcminutes, and an angular resolution of 6 arcseconds; typically the clusters in this sample have a radius of a few arcminutes. We set the center $\vec{x}_0$ of the cluster to be the center of mass, where pixel brightness is the mass equivalent. We then compute the $\mathcal{J}$-tensor given by Eq. \ref{eq:jfactortensor} across a rectangle of pixels centered around $\vec{x}_{0}$. 

In Fig. \ref{fig:axis_ratio_EGcluster}, we show the normalized distribution of the observed axis ratios in the cluster data, alongside the DM annihilation and decay signals expected for halos of masses larger than $2 \times 10^{12} M_\odot$ in order to increase statistics. We show the halos with the mass cut that matches that of the cluster data as arrows in blue (red) for annihilation (decay) in Fig. \ref{fig:axis_ratio_EGcluster}. Although with lower statistics, the sample with the same mass cut as the observed data as well as the extended sample show a tendency for axes ratios to extend to lower values that the observed data.

As we explain in App. \ref{sec:massdependence}, there is a trend for smaller halos to be more spherical, so we impose a mass cut on the Illustris halos to compare to the cluster data. However, note that the cluster sample may not be a representative sample of all clusters of similar masses. 
The observed clusters are quite symmetric about their centers of mass, and so it would be difficult to distinguish the astrophysical X-ray emission from a DM signal based on gross morphology alone, although the most spherical clusters (in the 0.9-1 bin) appear more symmetric than $98 \%$ ($60 \%$) of the halos studied in annihilation (decay) signals. In the same figure, we show the normalized distribution of the stars. The gas distribution was not included since it does not reproduce observational constraints in the case of clusters \cite{Genel:2014lma}. The X-ray data appears to be even more spherical than the star population, so indirect detection studies should not assume a spherical morphology for DM signals. More specific studies, such as analyzing possible signals using gravitational lensing, are required to understand extragalactic DM signals \cite{Graham:2015yga}.

\section{Conclusions}
\label{sec:conclusion}

In this paper, we studied morphological properties of DM Galactic and extragalactic indirect detection annihilation/decay signals, using the high-statistics Illustris simulation to map out the expected distribution of those properties. To understand the morphology of DM signals, we introduced two parametrizations for the asymmetry/elongation of an observed signal. The first is an analog of the inertia tensor called the $\mathcal{J}$-tensor; it weighs every pixel's contribution to the inertia tensor with the observed (DM) brightness. We also divided the sky into quadrants and studied the ratios of predicted signal brightness across opposite and adjacent quadrants. The advantage of these two methods is they are compatible with indirect detection observations.

We explored the DM signal morphology in two cases. In the first scenario, the observer is situated inside the halo, at a distance of 8.5 kpc from its center.
In this analysis we showed both results for the full halo sample, and for a subsample with DM mass and stellar mass comparable to the Milky Way. In this case, annihilation and decay signals are expected to be fairly symmetric, with the distribution of observed axis ratio peaking at $\sim0.9$. Halo substructure is more prominent in DM annihilation and the predicted signals appear slightly less symmetric compared to the case of decay, but these effects are minor. Baryons also play a role in making decay and annihilation distributions appear more spherical, but the effects are generally quite small, as the fraction of halos at any given axis ratio changes by a few percent.
We find that our results are fairly robust when the center or outer regions of the halos are excluded, and are only slightly affected by the presence of baryons.

In the second scenario we studied, the observer is external to the halo; this is the relevant analysis for searches for extragalactic DM signals. Both decay and annihilation signals are more frequently non-spherical than in the Galactic case; this is especially true for annihilation, where the distribution of axis ratio is very flat, and a sizable fraction of halos have a small axis ratio in the range $0.1 -0.5$. We believe that this tail can be largely attributed to halos possessing large subhalos, possibly due to recent halo mergers. Once halos with a substantial second subhalo are removed, the peak of the distribution shifts towards values of the axis ratio closer to 1. 

We examined the possible correlation between the baryonic disk and the principal axis of the decay/annihilation signal. 
We found that in the Galactic analysis, the signal's minor axis tended to be aligned with the angular momentum vector of the baryons, i.e. the direction perpendicular to the baryonic plane, although this correlation was quite mild (depending on the method of calculation, there were roughly $4-10\%$ more halos with $\theta < 0.1\pi$ than expected from the uncorrelated case, where $\theta$ is the angle between the DM signal's minor axis and the angular momentum vector). In the extragalactic analysis, we find a stronger correlation between the direction of the minor axis and the angular momentum vector of the baryons, as there is an excess of $\sim 62\%$ over the flat distribution for an angle $\theta < 0.1\pi$ between the halo's minor axis and the angular momentum vector. We think that the correlation is more pronounced in the extragalactic case first because the halos are largely non-spherical and therefore do have a preferred direction that does correlate with the baryons. This correlation is more pronounced for halos with a massive second subhalo. We think that this is due to the process of virialization; after the merger has occurred, the new subhalo is slowly getting virialized with the rest of the halo, and that process is sensitive to the presence of the baryons. 

Finally, we used two sets of observational data to study the degree to which DM signals might be distinguishable from astrophysical backgrounds: gamma-ray data from the \textit{Fermi} Gamma-Ray Space Telescope as an example for the Galactic analysis, and X-ray cluster data as a case study of potential extragalactic signals. The \textit{Fermi} all-sky data in the 2-12 GeV band have an axis ratio $\sim 0.5$, which is smaller than the axis ratio for Galactic decay signals from all tested MW-like halos (650 halos in 12 different orientations), and smaller than the axis ratio for Galactic annihilation signals more than $99\%$ of the time. When we remove estimates of the astrophysical backgrounds and examine the ``GeV excess'', focusing on the region around the Galactic center, we find that the residual is almost perfectly spherical, consistent with expectations for possible Galactic DM signals. Compared to the distributions of gas and stars, the background is closer to being part of the gas distribution, while the signal is more likely a DM signal. It is however difficult to exactly reproduce the MW morphology with the Illustris simulation. In contrast, the cluster X-ray maps are quite spherical, suggesting that it would be difficult to reliably exclude a DM origin for signals distributed similarly to the background, based on this approach alone.

To summarize, this study quantifies the degree of asymmetry to be expected in Galactic and extragalactic signals of DM annihilation or decay, putting the use of morphological data to separate potential signals from astrophysical background on a firmer footing.

\section*{Acknowledgements}
We thank Brendan Griffen, Federico Marinacci and Paul Torrey for their tremendous help in explaining the Illustris code and subsequent helpful conversations. We also thank Esra Bulbul for the X-ray images of clusters. We also thank the anonymous referee for helpful comments and feedback. We finally thank George Brova, Gabriel Collin, Alexander Ji, Ryan McKinnon, Ian Moult, Nicholas Rodd, Jesse Thaler, Mark Vogelsberger for helpful conversations. Some of the results in this paper have been derived using the HEALPix package~\cite{2005ApJ...622..759G}. Some computations in this paper were run on the Odyssey cluster supported by the FAS Division of Science, Research Computing Group at Harvard University. This work is supported by the U.S. Department of Energy under grant Contract Numbers DE-SC00012567 and DE-SC0013999. This research made use of the IDL Astronomy User's Library at Goddard. NB is supported by the S\~{a}o Paulo Research Foundation (FAPESP) under grants 2011/11973-4 and 2013/01792-8, by the Spanish MINECO under Grant FPA2014-54459-P and by the `Joint Excellence in Science and Humanities' (JESH) program of the Austrian Academy of Sciences.

\appendix 

\section{Analysis of Decaying Dark Matter}
\label{app:decay}
In the text, we have performed the quadrant analysis for the case of annihilating DM for an observer located at $R_\odot = 8.5~$ kpc as well as an observer situated well outside the halo. Here we perform a similar analysis for the case of decaying DM. 

\subsection{Galactic Analysis}

\begin{figure*}[t]
\begin{center}
\includegraphics[width=0.45\textwidth]{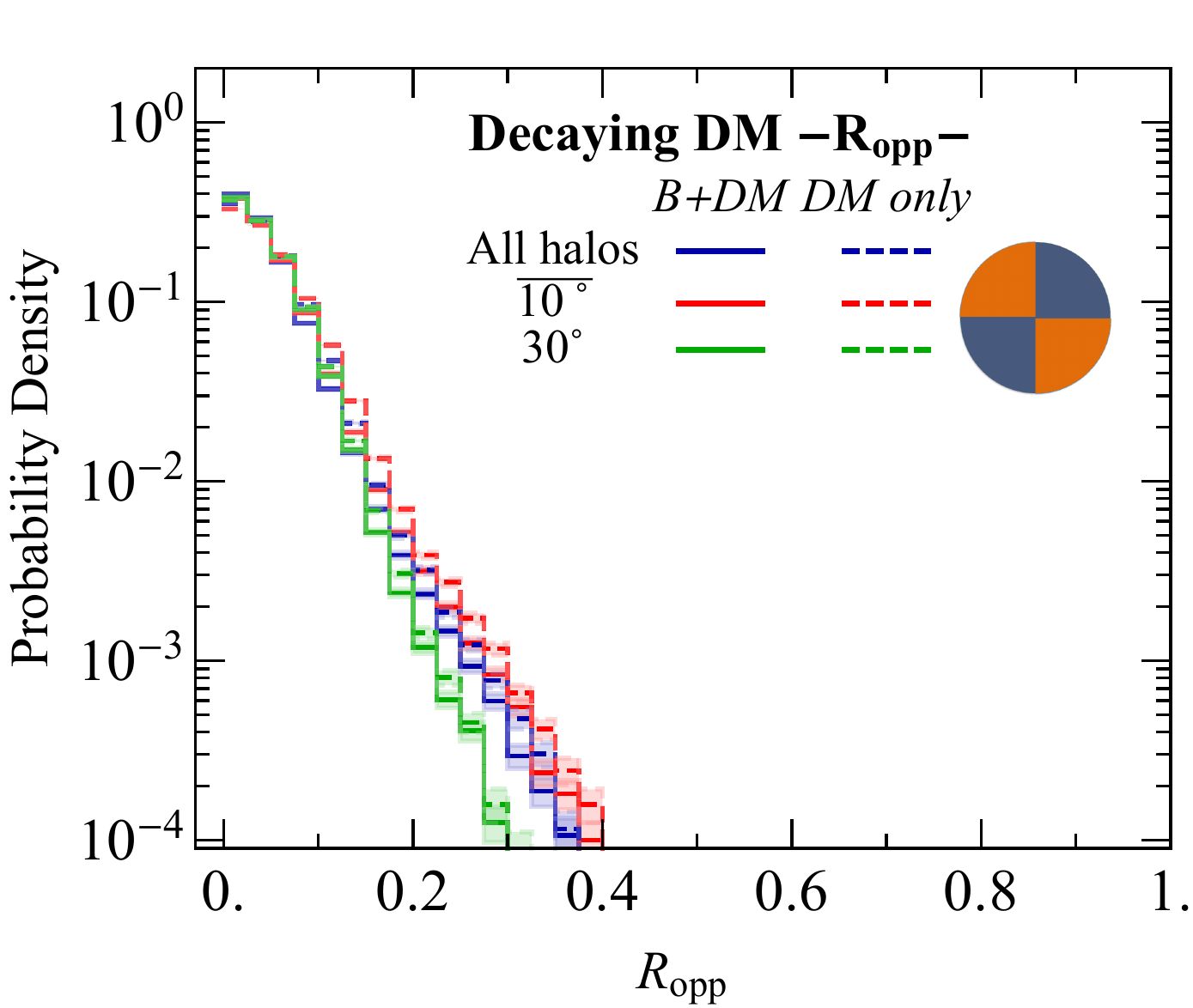}
\includegraphics[width=0.45\textwidth]{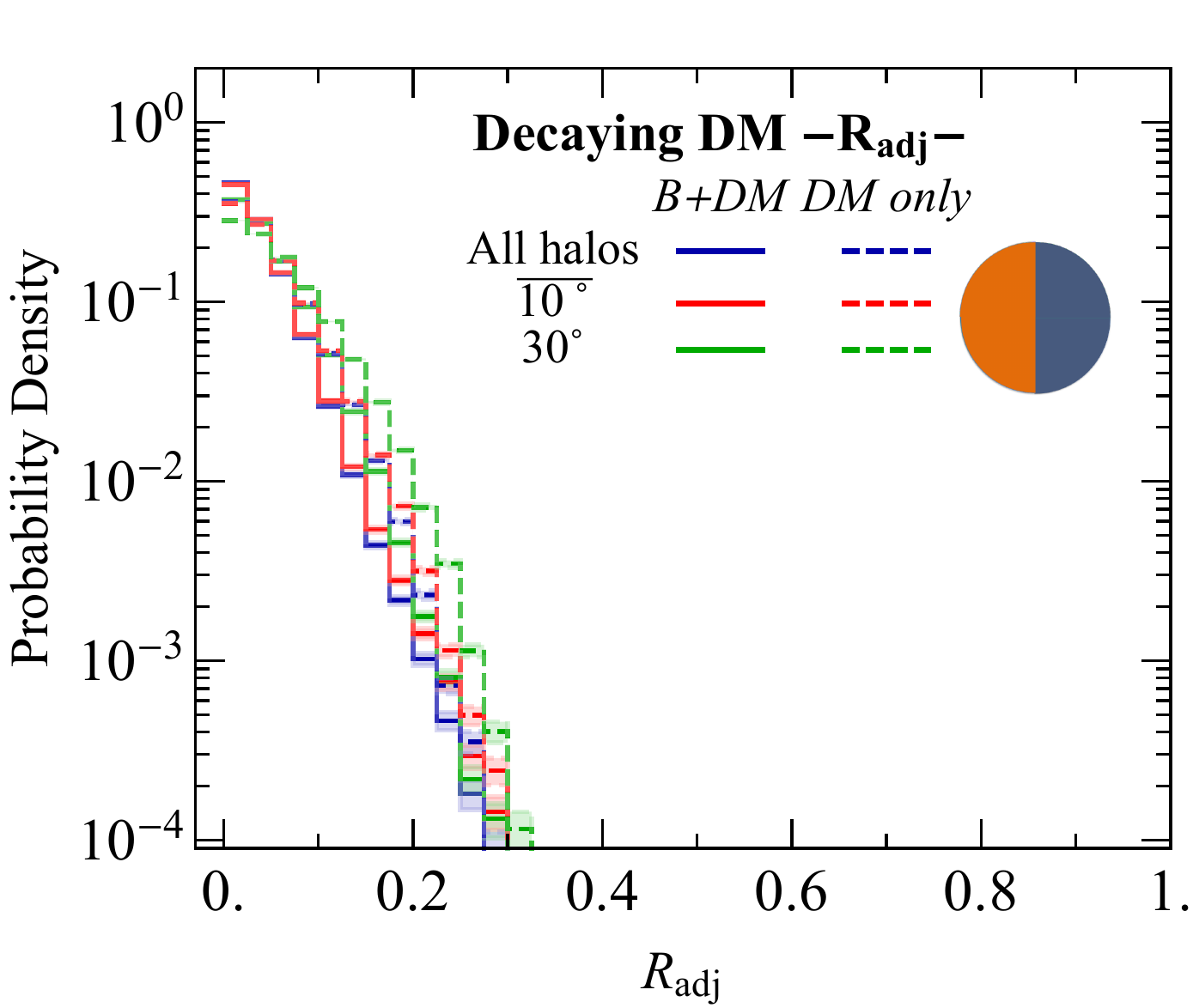}
\caption{\label{fig:illustrisdec}
As Fig. \ref{fig:illustrisann}, except for decay rather than annihilation.
}
\end{center}
\end{figure*}

\begin{figure*}[t]
\begin{center}
\includegraphics[width=0.45\textwidth]{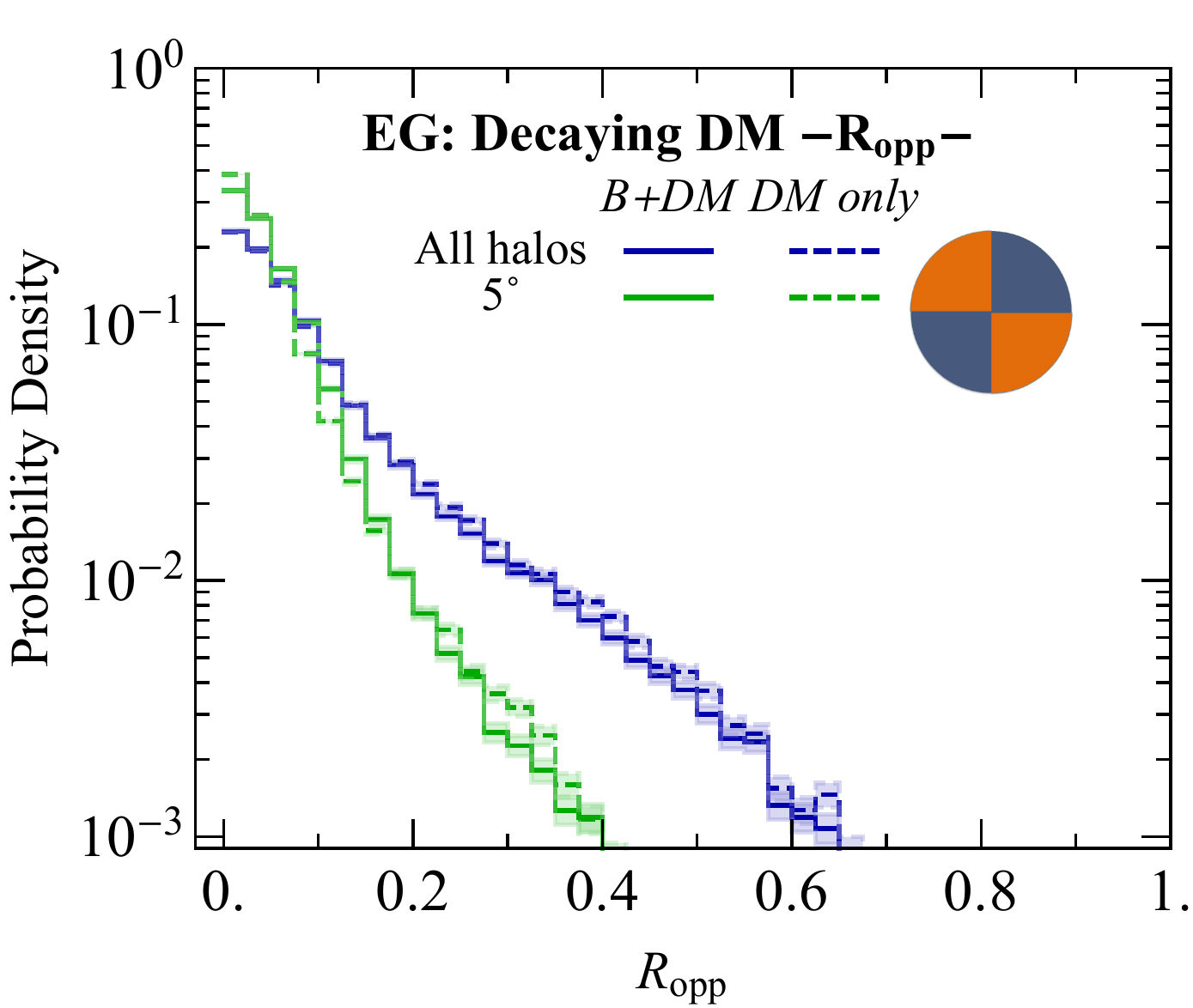}
\includegraphics[width=0.45\textwidth]{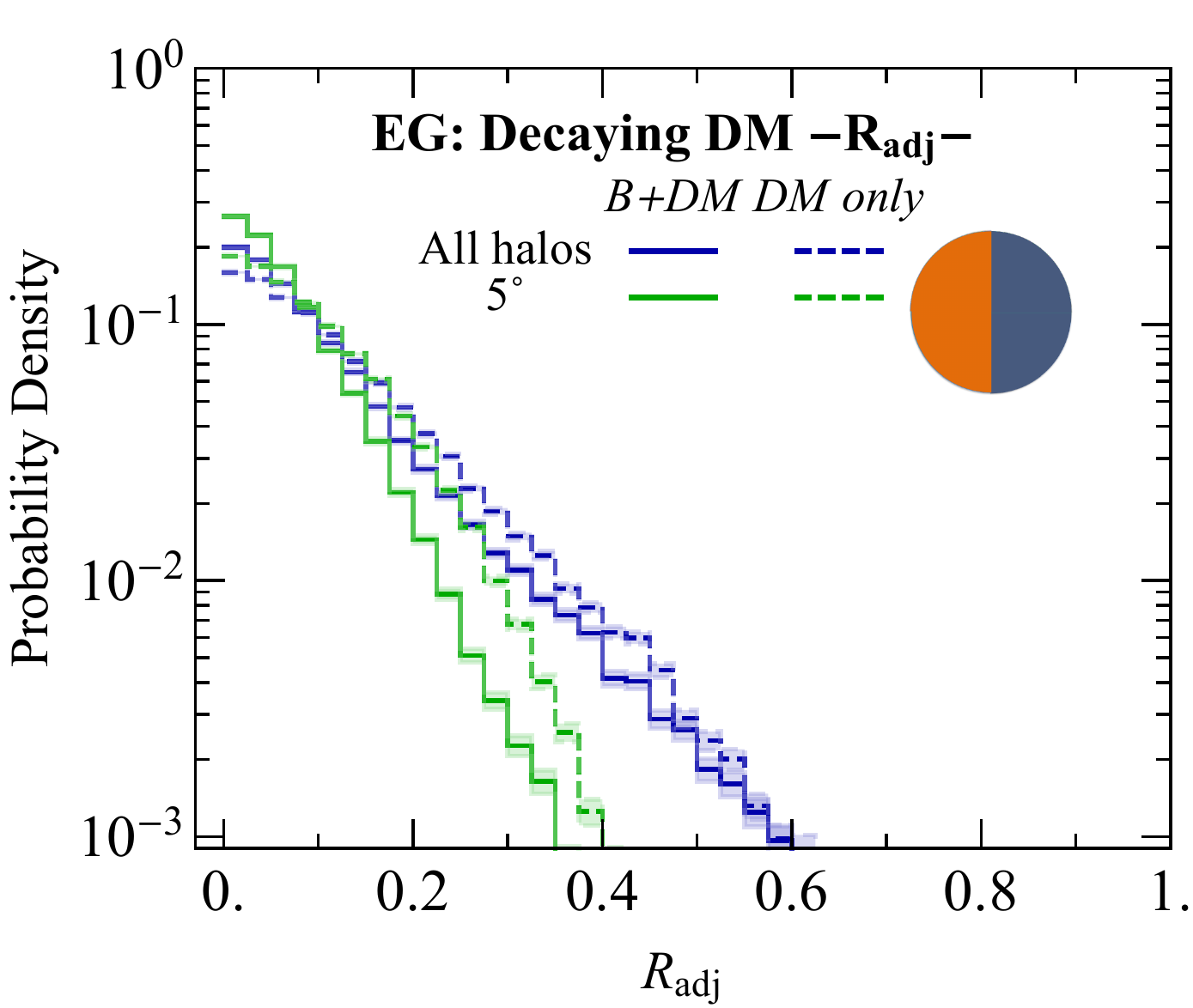}
\caption{\label{fig:illustrisEGdec}
As Fig. \ref{fig:illustrisEGann}, except for decay rather than annihilation.
}
\end{center}
\end{figure*}

Similarly to the analysis discussed in Sec. \ref{sec:results}, we show in Fig. \ref{fig:illustrisdec} the distribution of the variables $R_{\text{opp}}$ and $R_{\text{adj}}$ in the case of decaying DM. The differences between the three studied regions (the entire halo, the omitted inner cone of half angle 10$^\circ$, and the region up to $30^\circ$ from the center of the halo) are less pronounced as the $J$-factor in the case of decay compared to annihilation signals, but the results are consistent with the previous analysis of Sec. \ref{sec:IGallhalos}.

\subsection{Extragalactic Analysis}

As in Sec. \ref{sec:quad_EG}, We show in Fig. \ref{fig:illustrisEGdec} the probability distributions of $R_\text{opp}$ and $R_\text{adj}$ for the decay signals. The behavior is as expected from previous analyses; asphericity is less pronounced in decay signals compared to annihilation, and the data within a cone of half angle $5^\circ$ appears more symmetric. This is due to the off centered subhalos, as discussed in Sec. \ref{sec:mergers}.

\section{Mass Correlation}
\label{sec:massdependence}

The morphology of halos is highly mass-dependent \cite{Jing:2002np,Schneider:2011ed,2011MNRAS.413.1973W,2010MNRAS.407..581R}. We therefore categorize the masses of the halos of the Illustris simulation as follows:
\begin{itemize}
\item $M_{200} > 2 \times 10^{12} ~M_\odot$: This subset corresponds to the cluster-sized halos of the simulation.\footnote{We have increased this range of masses for larger statistics.} This subset is used to compare to the cluster X-ray data. 
\item $10^{10} ~M_\odot < M_{200} < 2 \times 10^{12} ~M_\odot$: This subset encompasses MW-way like halos as well as slightly less massive halos.
\item $M_{200} < 10^{10} ~M_\odot$: This is the subset for the least massive halos. 
\end{itemize}

\begin{figure*}[t]
\begin{center}
\includegraphics[width=0.45\textwidth]{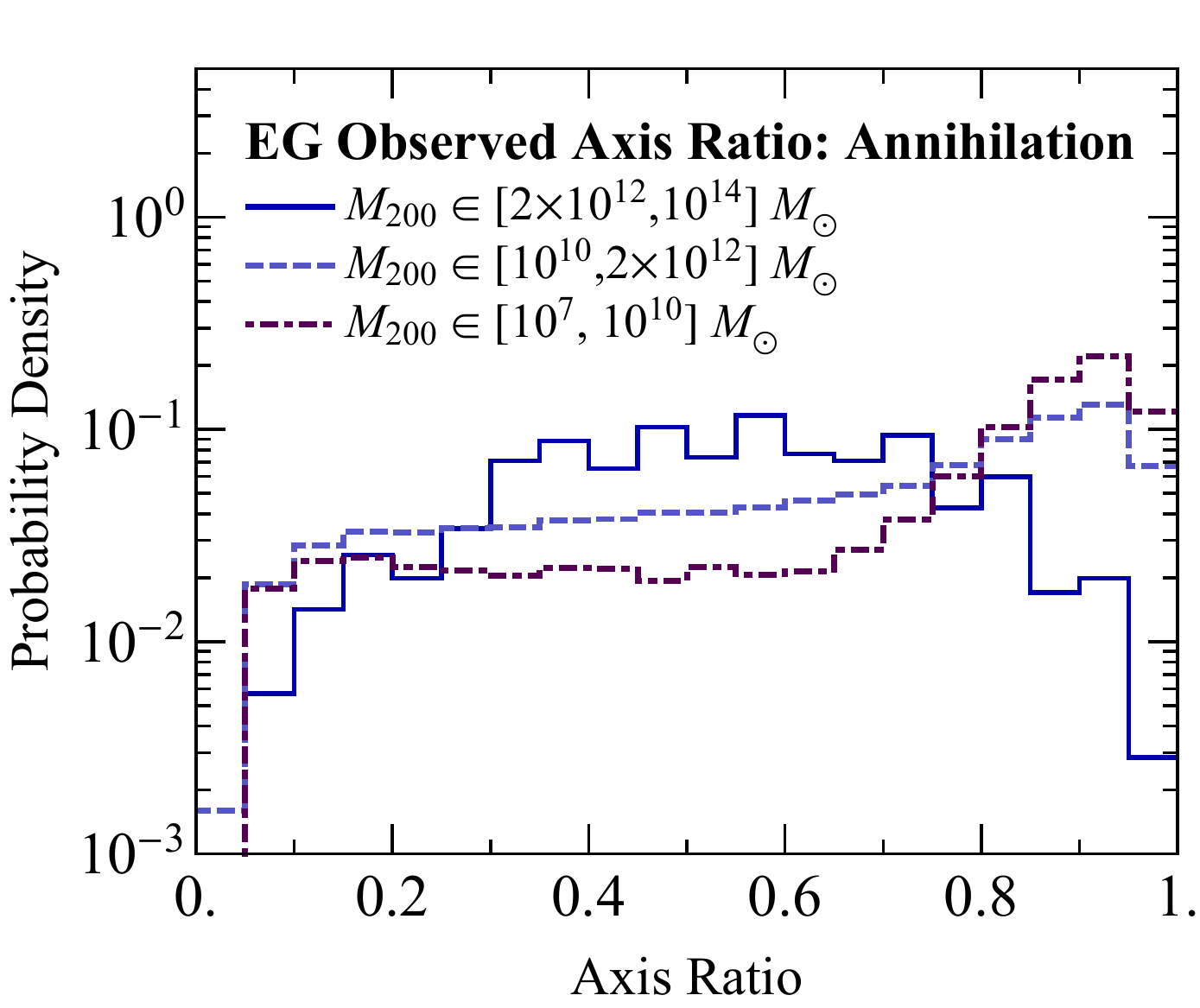}
\includegraphics[width=0.45\textwidth]{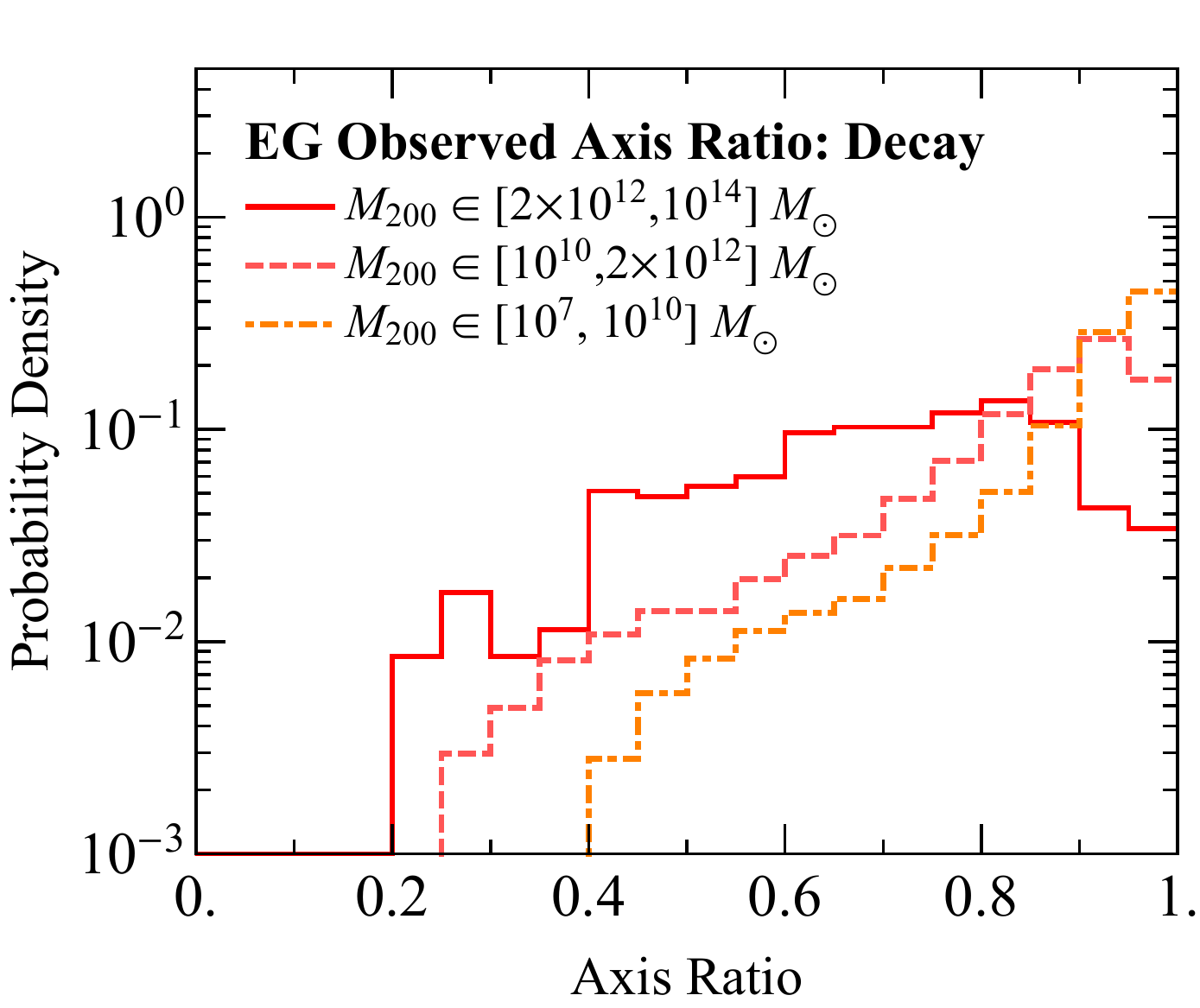}
\caption{\label{fig:illustrisEGmassann}
Histograms of the observed axis ratio for annihilation (left) and decay (right) for different mass bins: $ M_{200} > 2 \times 10^{12} M_\odot$, $10^{10} M_\odot < M_{200} < 2 \times 10^{12} M_\odot $ and $M_{200} < 10^{10} M_\odot$.
}
\end{center}
\end{figure*}

\begin{figure*}[t]
\begin{center}
\includegraphics[width=0.45\textwidth]{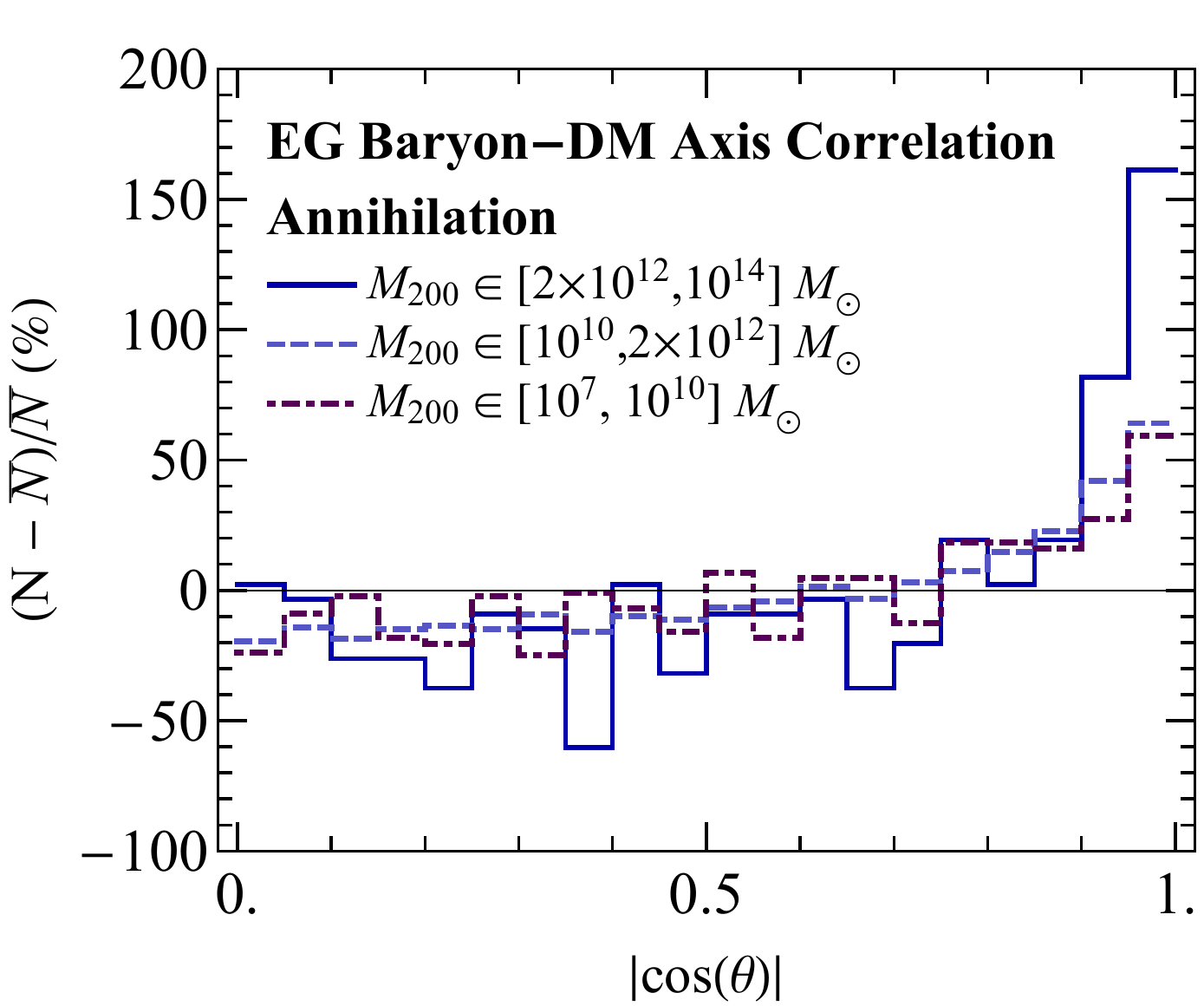}
\includegraphics[width=0.45\textwidth]{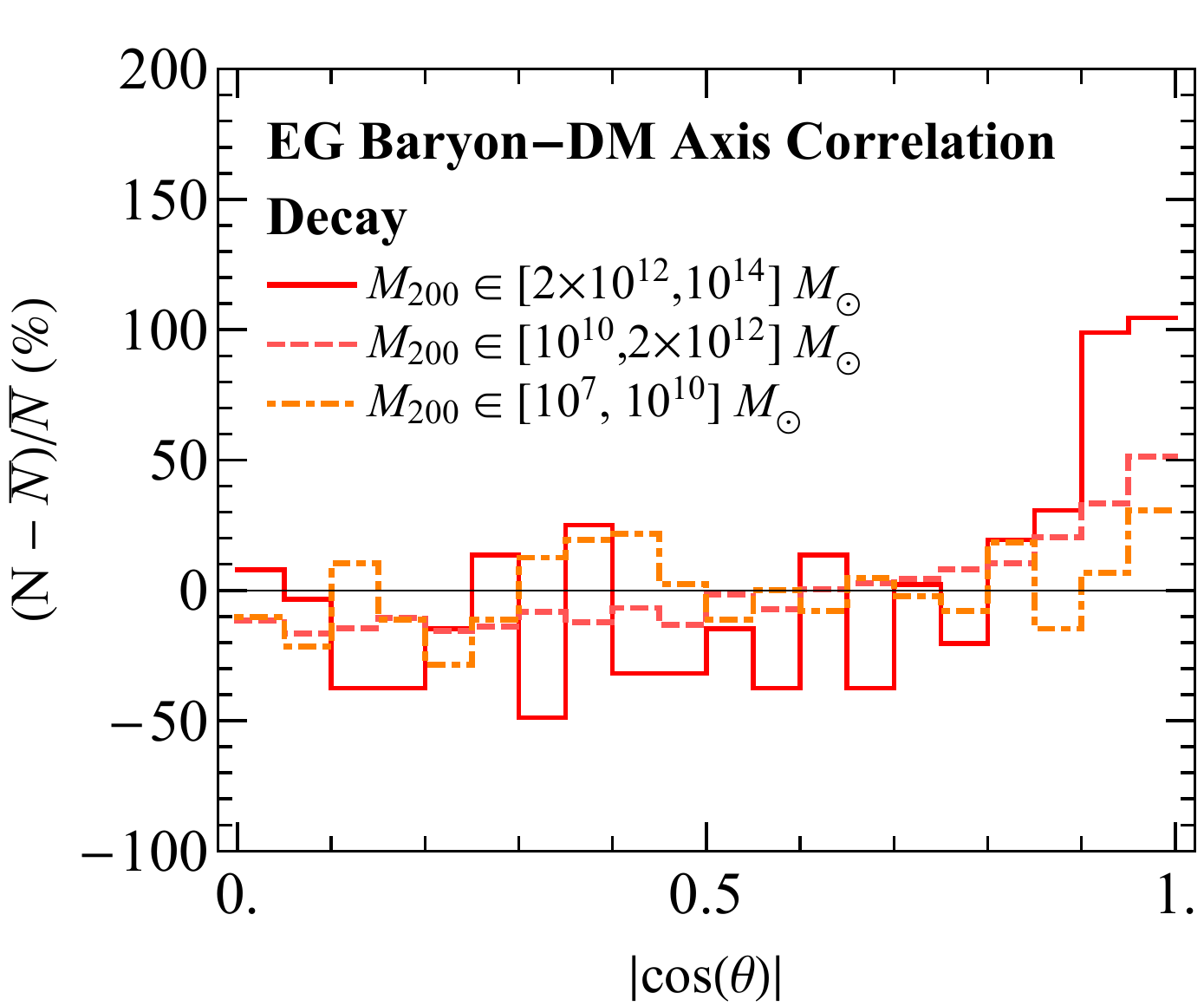}
\caption{\label{fig:angle_correleation_mass}
Histograms of the angle between the halo's minor axis and the angular momentum vector of the baryons in annihilation (left) and decay (right) for three different mass bins: $ M_{200} > 2 \times 10^{12} M_\odot$, $10^{10} M_\odot < M_{200} < 2 \times 10^{12} M_\odot $ and $M_{200} < 10^{10} M_\odot$.
}
\end{center}
\end{figure*}

In Fig. \ref{fig:illustrisEGmassann}, we plot the axis ratio for the different mass categories. We find consistent results that the more massive halos are the least expected to be spherical.
In Fig. \ref{fig:angle_correleation_mass}, we plot the angular correlation between the angular momentum vector and the halo's minor axis, as discussed in Sec. \ref{sec:anglecorrelation}, but now broken by mass category. We find that the most massive halos, which are the least spherical show indeed the most correlation with the baryonic axis.

\section{Comparison of Asymmetry Parameterization to Axis Ratio}

We define the following ratios as they relate to Eqs. \ref{eq:radj} and \ref{eq:ropp}.
\begin{align} \label{eq:ratios} r_\mathrm{opp} & = \frac{J_1 + J_3}{J_2 + J_4}, \quad & r_\mathrm{adj}  = \frac{J_1 + J_2}{J_3 + J_4}, \end{align}
and the related quantities
\begin{align} R_\mathrm{opp} & = \frac{| (J_1 + J_3) -( J_2 + J_4)|}{\sum_i J_i} = \frac{|r_\mathrm{opp} -1|}{r_\mathrm{opp}+1} \nonumber \\
R_\mathrm{adj} & = \frac{|(J_1 + J_2 )- (J_3 + J_4)|}{\sum_i J_i}  = \frac{|r_\mathrm{adj} -1|}{r_\mathrm{adj}+1}.   \end{align}

In the case that the annihilation/decay signal profiles are modeled as perfect ellipses (as done in e.g.~\cite{Daylan:2014rsa, Calore:2014xka}), the axis ratio is simply related to the parameter $r_\mathrm{opp}$, in the case where the quadrant boundaries lie at a $45^\circ$ degree angle to the major axis. 

Consider an arbitrary intensity function of the form $I(\sqrt{(x/a)^2 + (y/b)^2})$ (Note that real annihilation/decay profiles will not in general have this precise form).
Let us assume the signal is sufficiently localized that neglecting the curvature of celestial sphere is a reasonable approximation (which will be true if the halo is sufficiently distant, and for a peaked intensity profile is true even for our own Galaxy), so we can define the quadrant boundaries as simply $|x| = |y|$. 
Then:
\begin{align} r_\mathrm{opp} & = \frac{\int_{|x| > |y|} I(\sqrt{(x/a)^2 + (y/b)^2})\, dx\, dy}{\int_{|x| < |y|} I(\sqrt{(x/a)^2 + (y/b)^2})\, dx\, dy}. \end{align}

(There are two possible definitions of $r_\mathrm{opp}$ in this case, one of which is the reciprocal of the other. This is an arbitrary choice, so we expect the distributions of $r_\mathrm{opp}$ and $1/r_\mathrm{opp}$ to be identical. In this case, we will arbitrarily choose the quadrants in the numerator to be those lying along the $x$-axis.)

We can restrict ourselves to the region with $x > 0, ~y > 0$, and perform the integrals by the substitutions $X = x/a$, $Y = y/b$, followed by $X = R \cos \theta$, $Y = R \sin\theta$. Within this region, this procedure yields:

\begin{align}  & \int_{|x| > |y|} I(\sqrt{(x/a)^2 + (y/b)^2})\, dx\, dy \nonumber \\
& = a b \int_{|X| > (b/a) |Y|} I(\sqrt{X^2 + Y^2})\, dX\, dY \nonumber \\
& = a b \int_{0 < \tan\theta < a/b} I(R)\, R\, dR\, d\theta \nonumber \\
& = \tan^{-1}(a/b) \left[ a b \int dR\, I(R)\, R \right]. \end{align}
Similarly, 
\begin{align}  &  \int_{|x| < |y|} I(\sqrt{(x/a)^2 + (y/b)^2})\, dx\, dy \nonumber \\
& =  \left(\pi/2 - \tan^{-1}(a/b)\right) \left[ a b \int dR\, I(R)\, R \right]. 
\end{align}
Independent of the boundaries on the integral over $R$ or the details of the function $I(R)$, we thus obtain:
\begin{equation} r_\mathrm{opp} = \frac{\tan^{-1}(a/b)}{\pi/2 - \tan^{-1}(a/b)}, \quad R_\mathrm{opp} = \left|1 - \frac{4}{\pi} \tan^{-1}\frac{a}{b} \right| . \label{eq:axisratioconversion} \end{equation}
This result allows us to estimate limits on $r_\mathrm{opp}$ or $R_\mathrm{opp}$, when provided with limits on the axis ratio for a potential signal modeled as an ellipse, or vice versa.

In the limit where the ratio of major to minor axes is large (either $a/b \ll 1$ or $a/b \gg 1$), $r_\mathrm{opp}$ approaches $2 (a/b)/\pi$ if $a/b \ll 1$, and $\pi (a/b)/2 -1$ if $a/b \gg 1$. In the limit where $a/b \approx 1$, $r_\mathrm{opp} \approx 1 + \frac{4}{\pi} (a/b - 1)$, and $R_\mathrm{opp} \approx (2/\pi)|a/b - 1|$. Thus we may consider $r_\mathrm{opp}$ with a specific choice of quadrants as a rough proxy for axis ratio (while being more general, and well-defined for cases where the signal is not actually elliptical), with the approximation being most accurate for near-spherical intensity profiles. (Note that there will be corrections to Eq. \ref{eq:axisratioconversion} associated with the spherical coordinate system of the sky, and with any boundaries on the region of interest that are not only functions of $R$; if greater accuracy is desired in this conversion, $r_\mathrm{opp}$ should be computed numerically for the intensity profile and region of interest under study.)

Note that the parameter $r_\mathrm{adj}$ is identically 1 in the context of perfectly elliptical signal models, as the signal profile will be evenly bisected by any axis passing through its center.

\bibliography{sphericity}
\bibliographystyle{JHEP}

\end{document}